\documentclass[trackchanges, twocolumn]{aastex7} 

\usepackage{amsmath}	
\usepackage{amssymb}	
\usepackage{longtable}
\usepackage{scalerel}
\usepackage{hyperref}
\usepackage{mathtools} 

\usepackage{natbib}

\mathchardef\mhyphen="2D

\newcommand{\oiii}{O\,{\sc iii}}

\newcommand{\oviii}{O\,{\sc viii}}

\newcommand{\siiv}{Si\,{\sc iv}}

\newcommand{\cii}{[C\,{\sc ii}]}

\newcommand{\civ}{C\,{\sc iv}}
\newcommand{\mgii}{Mg\,{\sc ii}}
\newcommand{\mgxii}{Mg\,{\sc xii}}

\mathchardef\mhyphen="2D

\def\lya{Ly$\alpha$}
\def\ly{$\lambda$}
\def\ha{H$\alpha$}

\def\cii{C\,{\sc ii}}

\def\civ{C\,{\sc iv}}

\def\oi{O\,{\sc i}}
\def\oiii{O\,{\sc iii}}

\def\nex{Ne\,{\sc x}}

\def\mgii{Mg\,{\sc ii}}

\def\siiv{Si\,{\sc iv}}
\def\Siii{Si\,{\sc ii}}

\def\siiii{Si\,{\sc iii}}

\def\sii{S\,{\sc ii}}

\def\Q0059{Q0059--2735}

\def\S2S3{S2S3}

\definecolor{blk}{rgb}{0.0,0.0,0.0}
\definecolor{red}{rgb}{0.75,0.0,0.0}
\definecolor{yel}{rgb}{0.65,0.65,0.0}
\definecolor{grn}{rgb}{0.0,0.75,0.0}
\definecolor{blu}{rgb}{0.0,0.0,0.75}
\definecolor{gry}{rgb}{0.75,0.75,0.75}

\def\nh{\ifmmode n_\mathrm{\scriptscriptstyle H} \else $n_\mathrm{\scriptscriptstyle H}$\fi}
\def\ne{\ifmmode n_\mathrm{\scriptstyle e} \else $n_\mathrm{\scriptstyle e}$\fi}
\def\Te{\ifmmode T_\mathrm{\scriptstyle e} \else $T_\mathrm{\scriptstyle e}$\fi}
\def\Qh{\ifmmode Q_\mathrm{\scriptstyle H} \else $Q_\mathrm{\scriptstyle H}$\fi}
\def\Uh{\ifmmode U_\mathrm{\scriptstyle H} \else $U_\mathrm{\scriptstyle H}$\fi}
\def\Nh{\ifmmode N_\mathrm{\scriptstyle H} \else $N_\mathrm{\scriptstyle H}$\fi}
\def\Nhi{\ifmmode N_\mathrm{\scriptstyle HI} \else $N_\mathrm{\scriptstyle HI}$\fi}
\def\Uhhp{\ifmmode U_\mathrm{\scriptstyle H,HP} \else $U_\mathrm{\scriptstyle H,HP}$\fi}
\def\Nhhp{\ifmmode N_\mathrm{\scriptstyle H,HP} \else $N_\mathrm{\scriptstyle H,HP}$\fi}
\def\Uhvhp{\ifmmode U_\mathrm{\scriptstyle H,VHP} \else $U_\mathrm{\scriptstyle H,VHP}$\fi}
\def\Nhvhp{\ifmmode N_\mathrm{\scriptstyle H,VHP} \else $N_\mathrm{\scriptstyle H,VHP}$\fi}
\def\Nion{\ifmmode N_\mathrm{\scriptstyle ion} \else $N_\mathrm{\scriptstyle ion}$\fi}

\def\Zsun{\ifmmode {\rm Z}_{\odot} \else $Z_{\odot}$\fi}
\def\Msun{\ifmmode {\rm M}_{\odot} \else $M_{\odot}$\fi}
\def\kms{\ifmmode {\rm km~s}^{-1} \else km~s$^{-1}$\fi}
\def\Lya{\ifmmode {\rm Ly}\alpha \else Ly$\alpha$\fi}
\def\Lyb{\ifmmode {\rm Ly}\beta \else Ly$\beta$\fi}
\def\Lyg{\ifmmode {\rm Ly}\gamma \else Ly$\gamma$\fi}
\def\Lyd{\ifmmode {\rm Ly}\delta \else Ly$\delta$\fi}
\def\neaod{\ifmmode n_\mathrm{\scriptscriptstyle AOD} \else $n_\mathrm{\scriptscriptstyle AOD}$\fi}
\def\necrit{\ifmmode n_\mathrm{\scriptstyle cr} \else $n_\mathrm{\scriptstyle cr}$\fi}
\def\ncr{\ifmmode n_\mathrm{\scriptstyle cr} \else $n_\mathrm{\scriptstyle cr}$\fi}
\def\nepi{\ifmmode n_\mathrm{\scriptscriptstyle PI} \else $n_\mathrm{\scriptscriptstyle PI}$\fi}
\def\gtorder{\mathrel{\raise.3ex\hbox{$>$}\mkern-14mu\lower0.6ex\hbox{$\sim$}}}
\def\ltorder{\mathrel{\raise.3ex\hbox{$<$}\mkern-14mu\lower0.6ex\hbox{$\sim$}}}

\def\vro{\ifmmode v_\mathrm{\scriptscriptstyle 1, \scriptstyle r} \else $v_\mathrm{\scriptscriptstyle 1, \scriptstyle r}$\fi}
\def\vrc{\ifmmode v_\mathrm{\scriptscriptstyle 2, \scriptstyle r} \else $v_\mathrm{\scriptscriptstyle 2, \scriptstyle r}$\fi}
\def\vzo{\ifmmode v_\mathrm{\scriptscriptstyle 1, \scriptstyle z} \else $v_\mathrm{\scriptscriptstyle 1, \scriptstyle z}$\fi}
\def\vzc{\ifmmode v_\mathrm{\scriptscriptstyle 2, \scriptstyle z} \else $v_\mathrm{\scriptscriptstyle 2, \scriptstyle z}$\fi}

\newcommand{\Vcloud}{$v_\text{cl}$}
\newcommand{\Vwind}{$v_\text{wind}$}
\newcommand{\Vcir}{$v_\text{cir}$}

\newcommand{\HWHMcloud}{HWHM$_\text{cl}$}

\newcommand{\Mstar}{M$_{\star}$}

\newcommand{\Mdot}{$\dot{M}_\text{out}$}

\newcommand{\Mdotcloud}{$\dot{M}_\text{cool}$}
\newcommand{\Mdotcloudinit}{$\dot{M}_\text{cool,0}$}

\newcommand{\Mdotwindinit}{$\dot{M}_\text{hot,0}$}

\newcommand{\Pdotcloud}{$\dot{p}_\text{cool}$}

\newcommand{\EdotStar}{$\dot{E}_{\star}$}
\newcommand{\Edotcloud}{$\dot{E}_\text{cool}$}

\newcommand{\rhalf}{r$_{50}$}

\def\ZStar{\ifmmode {\rm Z}_\text{stars} \else $Z_\text{stars}$\fi}
\def\ZGas{\ifmmode {\rm Z}_\text{gas} \else $Z_\text{gas}$\fi}

\newcommand{\etam}{$\eta_\text{M}$}
\newcommand{\etamcool}{$\eta_\text{M,cool}$}
\newcommand{\etamhot}{$\eta_\text{M,hot}$}
\newcommand{\etaE}{$\eta_\text{E}$}
\newcommand{\etaehot}{$\eta_\text{E,hot}$}
\newcommand{\etaecool}{$\eta_\text{E,cool}$}
\newcommand{\etamcoolinit}{$\eta_\text{M,cool,0}$}
\newcommand{\etamhotinit}{$\eta_\text{M,hot,0}$}

\def\Nhtot{\ifmmode N_\mathrm{\scriptstyle H,\,tot} \else $N_\mathrm{\scriptstyle H,\,tot}$\fi}
\def\Nhcloud{\ifmmode N_\mathrm{\scriptstyle H,\,cl} \else $N_\mathrm{\scriptstyle H,\,cl}$\fi}
\def\ncloud{\ifmmode n_\mathrm{\scriptstyle H,\,cl} \else $n_\mathrm{\scriptstyle H,\,cl}$\fi}

\newcommand{\Rstar}{$r_{*}$}
\newcommand{\Rhalf}{$r_{50}$}

\newcommand{\Rcloud}{$R_\text{cl}$}
\newcommand{\Mcloud}{$M_\text{cl}$}
\newcommand{\Mcloudinit}{$M_\text{cl,0}$}
\newcommand{\Vturb}{$V_\text{turb}$}
\newcommand{\tcool}{$t_\text{cool}$}
\newcommand{\tmix}{$t_\text{mix}$}

\newcommand{\dNhdv}{$dN_\mathrm{\scriptstyle H}/dv$} 
 
\newcommand{\dNiondv}{$dN_\mathrm{\scriptstyle ion}/dv$} 

\newcommand{\chisq}{$\chi^{2}$}

\defcitealias{Fielding22}{FB22}

\revised{\today}

\begin{document}

\title{Resolving the Unresolved Galactic Winds in Multi-phase Models. I. \\ Methodology and Application}

\author[orcid=0000-0002-9217-7051]{Xinfeng Xu}
\affiliation{Department of Physics and Astronomy, Northwestern University,
2145 Sheridan Road, Evanston, IL, 60208, USA}
\affiliation{Center for Interdisciplinary Exploration and Research in
Astrophysics (CIERA), 1800 Sherman Avenue,
Evanston, IL, 60201, USA}
\email[]{xinfeng.xu@northwestern.edu}  

\author[orcid=0000-0003-3806-8548]{Drummond Fielding}
\affiliation{Department of Astronomy, Cornell University, 404 Space Sciences Building, Ithaca, NY 14853, USA}
\affiliation{Department of Physics, New York University, 726 Broadway Rm. 1005, New York, NY 10003, USA}
\email[]{drummondfielding@gmail.com}

\author[orcid=0000-0001-6670-6370]{Timothy Heckman}
\affiliation{Center for Astrophysical Sciences, Department of Physics \& Astronomy, Johns Hopkins University, Baltimore, MD 21218, USA}
\affiliation{School of Earth and Space Exploration, Arizona State University, Tempe, AZ 85287, USA}
\email[]{theckma1@jhu.edu}  

\author[orcid=0000-0003-2630-9228]{Greg L. Bryan}
\affiliation{Department of Astronomy, Columbia University, 550 West 120th Street, New York, NY 10027, USA}
\affiliation{Center for Computational Astrophysics, Flatiron Institute, 162 5th Avenue, New York, NY 10010, USA}
\email[]{gb2141@columbia.edu}  

\author[0000-0002-6586-4446]{Alaina Henry}
\affiliation{Space Telescope Science Institute, 3700 San Martin Drive, Baltimore, MD 21218, USA}
\email[]{ahenry@stsci.edu}  

\author[0000-0002-2644-3518]{Karla Z. Arellano-C\'ordova}
\affiliation{Institute for Astronomy, University of Edinburgh, Royal Observatory, Edinburgh, EH9 3HJ, UK}
\email[]{k.arellano@ed.ac.uk}

\author[0000-0003-4166-2855]{Cody Carr}
\affiliation{Center for Cosmology and Computational Astrophysics, Institute for Advanced Study in Physics, Zhejiang University, Hangzhou 310058, China}
\affiliation{Institute of Astronomy, School of Physics, Zhejiang University, Hangzhou 310058, China}
\email[]{carrx200@umn.edu}

\author[0000-0002-0302-2577]{John Chisholm}
\affiliation{Department of Astronomy, The University of Texas at Austin, 2515 Speedway, Stop C1400, Austin, TX 78712, USA}
\email[]{chisholm@austin.utexas.edu}

\author[0000-0001-6369-1636]{Claude-Andr\'e Faucher-Gigu\`ere}
\affiliation{Department of Physics and Astronomy, Northwestern University,
2145 Sheridan Road, Evanston, IL, 60208, USA}
\affiliation{Center for Interdisciplinary Exploration and Research in
Astrophysics (CIERA), 1800 Sherman Avenue,
Evanston, IL, 60201, USA}
\email[]{cgiguere@northwestern.edu}

\author[0000-0001-8587-218X]{Matthew Hayes}
\affiliation{Stockholm University, Department of Astronomy and Oskar Klein Centre for Cosmoparticle Physics, AlbaNova University Centre, SE-10691, Stockholm, Sweden}
\email[]{matthew.hayes@astro.su.se}

\author[0009-0002-9932-4461]{Mason Huberty}
\affiliation{Minnesota Institute for Astrophysics, University of Minnesota, 116 Church Street SE, Minneapolis, MN 55455, USA}
\email[]{huber458@umn.edu}

\author[0000-0002-3959-6572]{Michael Jennings}
\affiliation{Center for Astrophysical Sciences, Department of Physics \& Astronomy, Johns Hopkins University, Baltimore, MD 21218, USA}
\email[]{rjenni11@jhu.edu}

\author[0000-0001-9189-7818]{Crystal L. Martin}
\affiliation{Department of Physics, University of California, Santa Barbara, Santa Barbara, CA 93106, USA}
\email[]{cmartin@ucsb.edu}

\author[0000-0002-9136-8876]{Claudia Scarlata}
\affiliation{Minnesota Institute for Astrophysics, University of Minnesota, 116 Church Street SE, Minneapolis, MN 55455, USA}
\email[]{mscarlat@umn.edu}

\author[0000-0001-6369-1636]{Allison L. Strom}
\affiliation{Department of Physics and Astronomy, Northwestern University,
2145 Sheridan Road, Evanston, IL, 60208, USA}
\affiliation{Center for Interdisciplinary Exploration and Research in
Astrophysics (CIERA), 1800 Sherman Avenue,
Evanston, IL, 60201, USA}
\email[]{allison.strom@northwestern.edu}

\begin{abstract}

Galactic winds shape galaxy evolution; 
however, the outflowing gas is complex: it consists of multiple ionization phases, and its properties vary spatially.
Therefore, methods that combine high-fidelity observations with state-of-the-art galactic-wind models are limited. 
Here we investigate methods for fitting the column density profiles derived from high-quality outflow observations with the multiphase, multiscale wind model from \citet{Fielding22}. We identify three key outflow parameters: the initial hot-phase mass-loading factor (\etamhotinit), the initial cool-phase mass-loading factor (\etamcoolinit), and the initial cool-cloud mass (\Mcloudinit).
We obtain good fits (reduced $\chi^{2}<1.5$) for most galaxies, with tight constraints on \etamcoolinit\ and moderate constraints on the other two parameters.
We find the inferred \etamcoolinit\ and \etamhotinit\ are mostly of order unity, with significant scatter. The constraints on \etamhotinit\ suggests that the interaction between the cool and hot phases allow us to indirectly constrain the properties of the hot wind from cool-outflow observations.
The model also predicts various radial trends. First, for all galaxies, the cool-phase outflow velocity increases rapidly between $1-2$\Rhalf, then reaches a plateau, where \Rhalf\ is the half-light-radius of the galaxy. Second, most galaxies exhibit increasing \etamcool\ and decreasing \etamhot\ with radius, with a few showing the reverse trends. These results should be interpreted as effective, model-conditional constraints but are consistent with other recent multiphase simulations and observations. This highlights that the velocity-radius mapping encoded in UV absorption profiles enables recovery of outflow spatial structures from spatially-integrated spectra. Our method paves the way for future broad parameter studies and guides updates of outflow simulations in future work.

\end{abstract}


\section{Introduction} 
\label{sec:intro}
\setcounter{footnote}{0} 

The evolution of galaxies is not a one-way street: while baryonic matter collapses to form stars and planets, this growth is regulated by feedback from stellar winds, supernovae (SNe), and active galactic nuclei (AGN), which return mass, energy, and momentum to the interstellar medium (ISM) and circumgalactic medium (CGM). Understanding how these processes are linked remains one of the central scientific goals of the decade \citep[e.g.,][]{Naab17, FG23}.

Galactic-scale winds represent one of the major phenomena responsible for delivering these feedback effects. It is well established that galactic winds consist of multiple ionization phases, with temperatures spanning from $\sim$ 10 K to $\gtrsim$ 10$^{7}$ K \citep[see reviews in][]{Veilleux20, Thompson24}. For galactic winds driven by star formation (SF)—the primary focus of this work—the widely accepted picture suggests that the volume-filling, hot wind fluid (T $\gtrsim$ 10$^{7}$ K) is powered by the output from young, massive stars, including stellar winds, radiation, and SNe. This hot wind fluid subsequently interacts with ambient cooler clouds (T $\sim$ 10–10$^{4}$ K) in the ISM, accelerating them outward at velocities up to $\sim$ 10$^{3}$ km s$^{-1}$, as well as increasing or decreasing their masses due to mixing or radiative cooling. Nevertheless, it remains difficult to observe outflow signatures simultaneously across all phases--with the exception of a few nearby sources, e.g., for the prototypical starburst M~82 \citep[see][]{Strickland09, Leroy15, Martini18, Xu23c, Boettcher24, Fisher25, Lopez25}. 

Given the greater accessibility of rest-frame UV and optical spectra for galaxies, most current observations of galactic winds have concentrated on detecting the cool outflowing clouds with T $\sim$ 10$^{4}$ K (hereafter, outflow clouds specifically refer to this component). Common diagnostic lines tracing these outflows include \lya\ \ly 1216, \Siii\ multiplet (e.g., 1260, 1304, 1526), \siiv\ \ly\ly 1393, 1402, \civ\ \ly\ly 1548, 1550, and \mgii\ \ly\ly 2796, 2803 in the UV, as well as [\oiii] \ly\ly 4959, 5007, \ha\ \ly 6564, and [\sii] \ly\ly 6718, 6732 in the optical \citep[e.g.,][]{Marasco23, Martin24, deGraaff24, Peng24, SaldanaLopez25, Shaban25, Xu25}. On the contrary, while the hot wind fluid is expected to carry the majority of the energy budget of the wind \citep[e.g.,][]{Li+Bryan20}, direct constraints on it are only available for limited nearby starburst galaxies, primarily through detections of X-ray He-like Fe line emission \citep[e.g.,][]{Heckman90, Lehnert96, Mitsuishi11, Wang14, Lopez20}. Thus, developing methods to constrain the hot phase through observations of the cooler components would represent a powerful advance toward understanding how galactic winds regulate galaxy growth. Given the expected transfer of mass, momentum, and energy between the hot and cool phases as they make their way out of the galaxy \citep[e.g.,][]{Ji19, Gronke20, Fielding22}, this could be possible, though it has not yet been validated in the literature.

In addition to the complexity of galactic winds' multiphase structure, they also span a wide range of spatial scales—from large-scale propagation across the entire galaxy to the fundamentally small-scale interactions between the galaxy’s ISM/CGM and the wind material. This broad spatial range further complicates both the observation and theoretical modeling of galactic winds. In observations, spatially resolved spectroscopic detections of outflows are costly and technically demanding, resulting in only a limited number of dedicated studies in the literature \citep[e.g.,][]{Burchett21, Zabl21, Shaban22, Shaban23, Xu23c, Lamperti24, Reichardt24, Rodriguez24, Parlanti25}. In models, it remains challenging to simultaneously capture both small and large scales under realistic settings. Specifically, properly resolving the small-scale interaction between the ISM/CGM and wind material requires parsec-scale resolution, which limits simulations to relatively small computational domains \citep[cosmologically speaking, see, e.g.,][]{Smith18, Hu19, Gronke20, kim20, Schneider20}. 

Overall, the multiphase, multiscale nature of galactic winds poses major challenges to both observations and simulations, especially when the two must be compared directly.
Here we tackle these challenges by combining high-quality observations with state-of-the-art analytic wind models that account for mixing between the hot and cool phases (see details in Section~\ref{sec:data}).
In this Paper~I, we ask whether this approach works and quantify how well the model parameters are constrained.
In a future Paper~II, we will broaden the parameter exploration, refine the model hierarchy, and calibrate the preferred configurations against the observations.

The structure of this paper is as follows. In Section \ref{sec:data}, we describe the galaxy sample and multiphase, multiscale galactic wind model we adopt. We present our main results in Sections \ref{sec:single} and \ref{sec:whole}, including how we constrain outflow properties from the hot and cool phases for our galaxies. We discuss comparisons with previous models and observations, caveats, and future plans in Section \ref{sec:discuss}. We conclude the paper in Section \ref{sec:conclusion}.

\begin{figure*}
\center
	\includegraphics[page = 1, angle=0,trim={0.0cm 0.0cm 0cm 0.0cm},clip=true,width=1.0\linewidth,keepaspectratio]{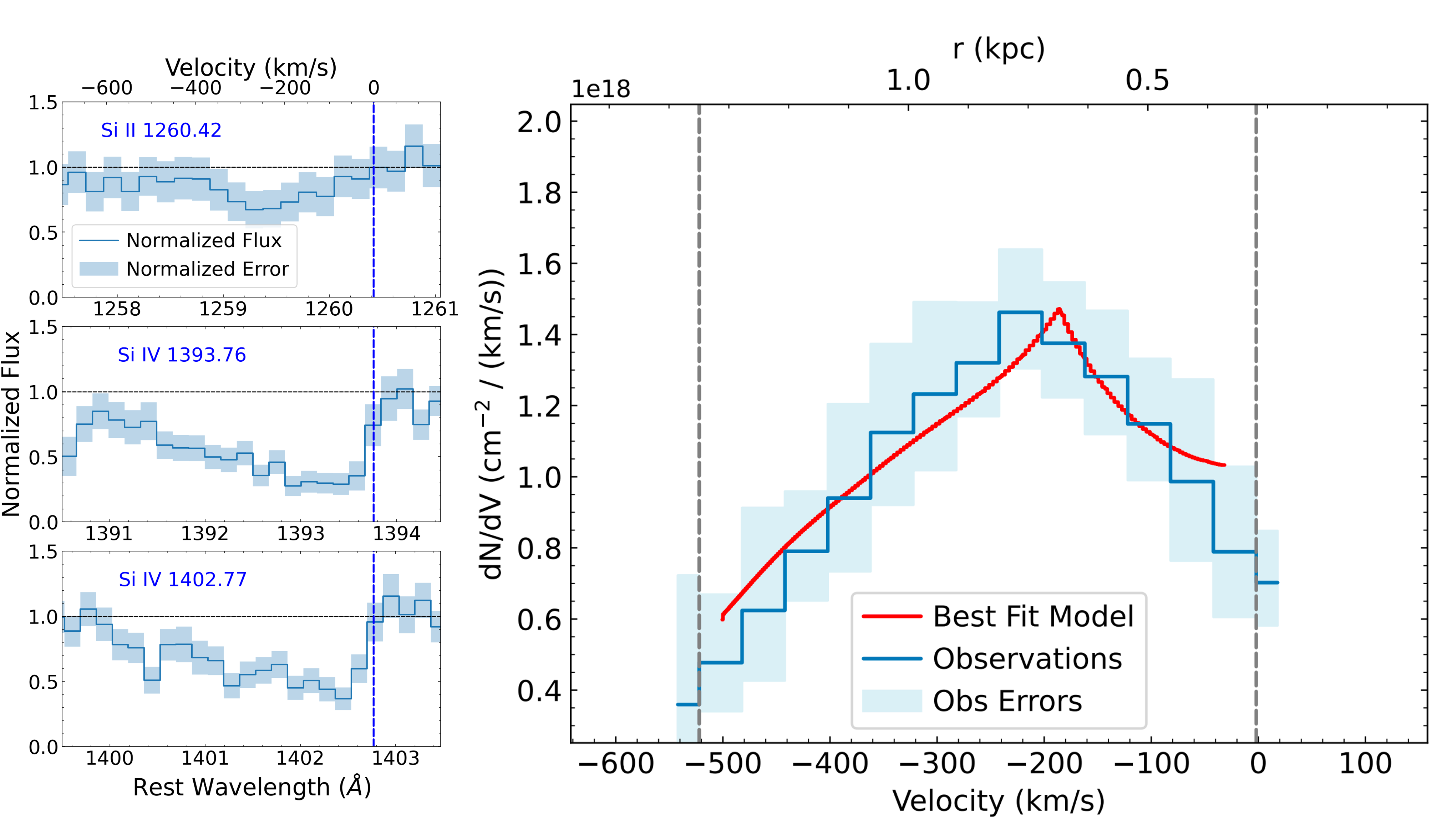}

\caption{\normalfont{An example of observations and the best-fit model for galaxy J0021+0052 (z = 0.0983). \textbf{Left:} HST/COS rest-UV spectra for \Siii\ and \siiv\ lines \citep{Xu22a}. Significant blue-shifted absorption lines denote the detections of galactic outflows. The vertical blue dashed line represent the systemic redshift of the galaxy (determined from optical emission lines). \textbf{Right:} Comparisons of the measured dN/dV profile in blue and the best-fit \citetalias{Fielding22} model in red. The profiles between the two gray lines are used in the fit. In the top axis, we show the model predicted radius of the outflow clouds in unit of kilo-parsec. See details of the model settings in Section \ref{sec:param}, and discussion of fit results in Section \ref{sec:single}.} }
\label{fig:single_model}
\end{figure*}

\section{Methodology} 
\label{sec:data}

In this Section, we describe the observational dataset in Section~\ref{sec:CLASSY}, and the wind model in Section~\ref{sec:FBModel}. In Section~\ref{sec:param}, we present our methodology to fit the observations with the model.

\subsection{Observations and Outflow Measurements} 
\label{sec:CLASSY}
Our sample includes 22 local star-forming galaxies jointly compiled from the COS Legacy Archive Spectroscopy SurveY (CLASSY) atlas \citep[][]{Berg22, James22} and the sample presented by \citet{Heckman15}. These galaxies were observed with the Hubble Space Telescope (HST)/Cosmic Origins Spectrograph (COS), forming the largest and highest-quality UV spectroscopic dataset of local star-forming galaxies. These objects span a wide range of important galaxy properties, including  stellar mass (log \Mstar\ $\sim6-11\ M_\odot$), SFR ($\sim 0.01 - 100$ M$_\odot$ yr$^{-1}$), metallicity (12+log(O/H)$\sim7-9$), and [\sii]-based electron density ($n_e\sim10^1-10^3$ cm$^{-3}$). The COS observations, given an aperture\footnote{COS aperture is affected by significant vignetting at radii greater than 0.4\arcsec.} of 2.5\arcsec, result in spatially integrated spectra covering the far-ultraviolet (FUV) wavelength range from $\sim$ 1200 to 2000\AA. To improve the signal-to-noise ratio (SNR), the spectra are resampled into bins of 0.18\AA, corresponding to a resolving power of $R\sim6000$–$10000$, or velocity resolution of 50–30 km s$^{-1}$ from the blue to red ends. Further details on the program design and data reduction can be found in previous publications \citep[][]{Berg22, James22, Heckman15}.

Given the dedicated UV dataset, cool outflow properties in these galaxies have been studied in various publications \cite[e.g.,][]{Xu22a,Xu23a,Hu23, Huberty24, Parker24}. We take the one that focus on the rest-UV absorption lines \citep{Xu22a}. In brief, outflow signatures are detected in lines including \oi\ \ly 1302, \cii\ \ly 1334, and \Siii\ multiplet (\ly1190, 1193, 1260, 1304, and 1526), and from higher-ionization transitions, e.g., \siiii\ \ly 1206, and \siiv\ \ly\ly 1393, 1402 (see examples in the left panel of Figure \ref{fig:single_model}). Then velocity-dependent ionic column density (\dNiondv) is measured using the partial coverage method \citep[PCM, e.g.,][]{Rupke05, Martin09, Chisholm16a, Rivera-Thorsen17}. These column densities are then fed into the CLOUDY photoionization models \citep[v17.01][]{Ferland17} to constrain the total hydrogen column density per velocity (\dNhdv). An example of the \dNhdv\ profile is shown as the blue lines in the right panel of Figure \ref{fig:single_model}. After that, \cite{Xu22a} adopts spherical outflow models to estimate the average mass, momentum, and energy outflow rates (\Mdotcloud, \Pdotcloud, \Edotcloud), which represent the spatially integrated values within the COS aperture. Finally, the cool outflows' mass loading factors (\etamcool) are calculated as \Mdotcloud/SFR, where SFR is derived from spectral energy distribution (SED) fittings described in \cite{Berg22}.

\subsection{Multiphase, Multiscale Wind Models}
\label{sec:FBModel}

To model galactic winds, one must capture both their multiphase and multiscale character—that is, how distinct temperature phases interact as the flows propagate from the galaxy into the CGM. In addition, to compare with observational results (Section~\ref{sec:CLASSY}), the model must remain flexible enough to span diverse galaxy properties, including SFR, $M_\star$, and the outflow mass loading factors in different phases ($\eta$). 

With these requirements in mind, we adopt the state-of-art analytic wind model developed by \citet{Fielding22} (hereafter \citetalias{Fielding22}). In this framework, the galactic wind consists of a hot, volume-filling component (the hot wind) and a cool component composed of embedded outflowing clouds (the cool outflows). As these two phases coevolve, they exchange mass, momentum, and energy through turbulent radiative mixing layers (TRMLs), a process motivated by results from high-resolution simulations \citep[e.g.,][]{Ji19, Gronke20, Mandelker20, Abruzzo24, Marin-Gilabert25}. The fate of the cool clouds is governed by the relative timescales of the radiative cooling (\tcool) in TRMLs surrounding each cloud and their mixing with the hot wind (\tmix\ = \Rcloud/\Vturb), where \Vturb\ is the turbulent velocity. Large clouds with \tcool\ $<$ \tmix\ are able to condense additional mass from the hot wind and grow. In contrast, smaller clouds with \tcool\ $>$ \tmix\ are shredded by turbulent mixing with the hot wind. This ongoing mass and energy exchange not only determines the survival of the cool clouds, but also modifies the global structure of the hot wind relative to classical single-phase models. 

The model considers that mass and energy are injected uniformly inside the star-forming region radius (r $<$ \Rstar), and cooling and gravity are not included within this region.
Beyond this radius, radiative cooling and gravity are included, which shape the coupled evolution of the wind and clouds as they traverse the surrounding medium. 
Cooling curves from \citet{Wiersma09} are adopted, including heating from the z=0 UV background.
The flow is treated as one-dimensional, with forces confined to the radial direction.
In addition, we assume a spherically symmetric galactic wind in the model, which is compatible with the large outflow solid angle ($\Omega$) inferred from our observations \citep{Xu22a}.  The temperature of the cool outflow clouds are set to be T = 10$^4$ K, which also matches the one taken for the observed UV absorption lines. We assume the hot wind has a metallicity of 2$Z_\odot$ \citep[e.g.,][]{Lopez20}, while we adopt the observed gas-phase metallicity for the cool outflows \citep{Berg22}.

\subsection{Methodology to Fit Observations with Models}
\label{sec:param}

In what follows, we show how the multiphase, multiscale \citetalias{Fielding22} framework is applied to fit single-phase, spatially integrated outflow measurements. The first task is to select observables that capture the properties of the cool outflow clouds and, via these, indirectly constrain the hot wind and the outflow’s radial structure.

To characterize the cool outflow clouds, the key observables are their mass, momentum, and energy outflow rates, which track the strength and impact of feedback.
We start with the observational analysis of outflows. Under a spherical thin-shell outflow model, mean rates follow directly from the velocity–resolved column density profile \citep[e.g.,][]{Xu22a}:

\begin{equation}\label{eq:Mdot}
    \begin{aligned}
        <\dot{M}_\text{cool}> &= \Omega \mu m_p R_0 \int \frac{dN_\text{H}}{dv}\times vdv\\
        <\dot{p}_\text{cool}> &= \int \dot{M}_\text{cool}(v) v \\
            &= \Omega \mu m_p R_0 \int \frac{dN_\text{H}}{dv}\times v^{2}dv\\
       <\dot{E}_\text{cool}> &= \int \frac{1}{2}\dot{M}_\text{cool}(v) v^{2}\\
            &= \frac{1}{2}\Omega \mu m_p R_0 \int \frac{dN_\text{H}}{dv}\times v^{3}dv\\
    \end{aligned}
\end{equation}
where $\mu$ is the mean atomic mass per proton ($\sim 1.4$), $m_p$ is the proton mass, $R_{0}$ is the mean outflow radius from which the outflow begins, $v$ is the mean outflow velocity, and $\Omega$ is the global covering factor, which approaches $4\pi$ as described in \citet{Xu22a}. We adopt $R_0 = 2 \times r_{50}$,  where $r_{50}$ is the UV half-light radius of the star-forming region. This choice is consistent with previous studies \citep[e.g.,][]{Shopbell98, Heckman15, Xu23a} and corresponds to an outflow that is launched at a radius enclosing $\sim$ 90\% of the starburst. The remaining free parameter is the velocity-dependent total hydrogen column density of the clouds (\Nh$(v)$).

Given the moderate spectral resolution of HST/COS (30–50 km s$^{-1}$ in our case), our galaxies enable velocity-resolved measurements of \Nh$(v)$. An example of it appears as the blue line in the right panel of Figure \ref{fig:single_model}. 
We fit three summary quantities of \Nh$(v)$ that capture its normalization, characteristic velocity, and width of the profile.
This choice balances observability and fidelity: each quantity is straightforward to measure and together they reproduce the profile’s overall shape.
Specifically, we use (1) the integrated hydrogen column of the outflowing clouds (\Nhcloud), obtained by integrating over $v < 0$; 2) the bulk velocity, \Vcloud, defined as the velocity at which \dNhdv\ reaches its maximum; 3) the half width at half maximum on the blueshifted side (\HWHMcloud), i.e., the width of the high-speed wing with $v <$ \Vcloud.
We do not use the full width because the \citetalias{Fielding22} model produces only outflows, so the modeled profiles occupy only the blueshifted side ($v < 0$). 

\citetalias{Fielding22} model directly yields the velocity and number density of outflow clouds at each radius (\Vcloud$(r)$ and \ncloud$(r)$). From that, we have:

\begin{equation}\label{eq:dNdv}
    \frac{dN_\text{H,cl}}{dv}(r)
    = n_\text{H,cl}(r) \left/ \frac{dv_{cl}(r)}{dr} \right.
\end{equation}
We then measure \Nhcloud, \Vcloud, \HWHMcloud\ from this modeled \dNhdv\ profile and fit them to the ones measured from observations.

\begin{figure*}
\center
	\includegraphics[page = 1, angle=0,trim={0.0cm 0.0cm 0cm 32.0cm},clip=true,width=1.0\linewidth,keepaspectratio]{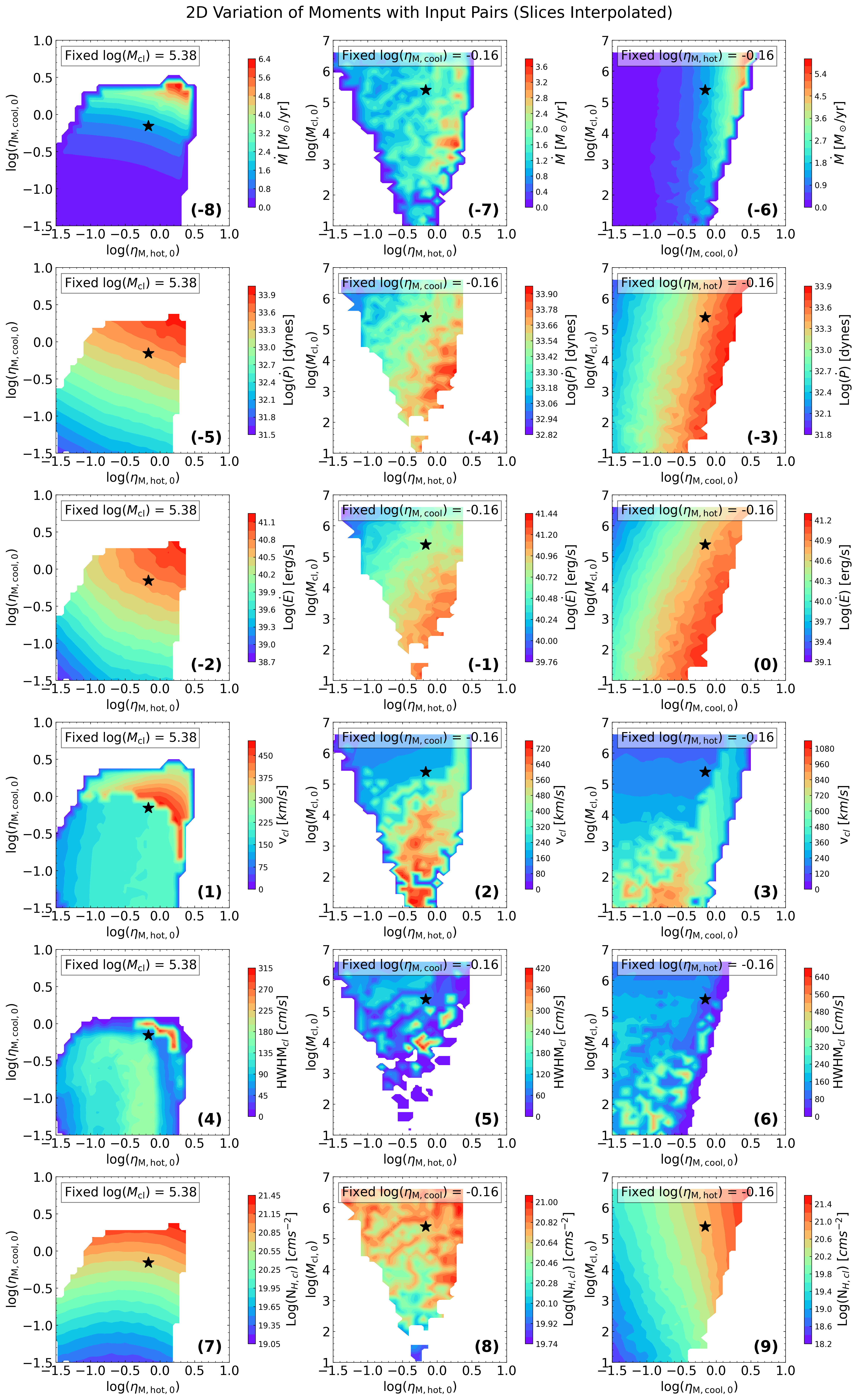}
    
	\includegraphics[page = 1, angle=0,trim={0.0cm 0.0cm 0cm 0.0cm},clip=true,width=1.0\linewidth,keepaspectratio]{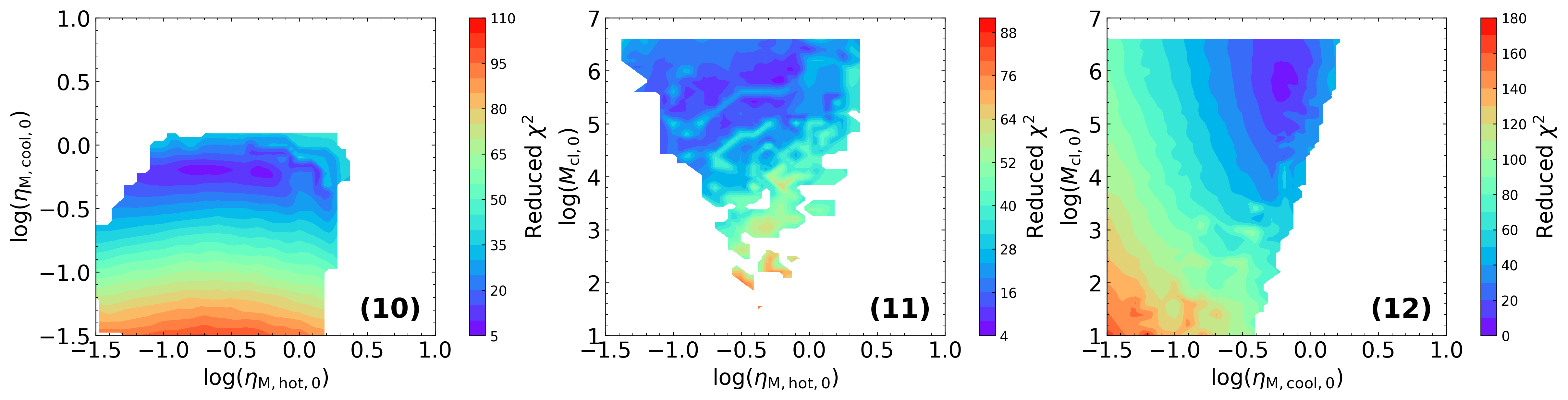}
    
\caption{\normalfont{Three-dimensional grid models illustrating the parameter space for J0021+0052. From panels (1) to (9), we vary two of the three input parameters (\etamhotinit, \etamcoolinit, \Mcloudinit), while keeping the third fixed at its best-fit value for J0021+0052 (see Figure~\ref{fig:single_model}). Therefore, each panel can be treated as a slice in the three-dimensional grids. In the first three rows, we draw color scales using the three summary quantities of the cool-phase outflows, including the bulk outflow velocity (\Vcloud), half-width-half-maximum (\HWHMcloud), and the integrated hydrogen column (\Nhcloud). More details can be found in Section \ref{sec:param}. Black stars denote the best-fit model's location, while the color gradients indicate parameter sensitivity. For example, in panel (7), the nearly horizontal color strips indicate that \Nhcloud\ strongly depends on \etamcool\ but very weakly on \etamhot. In the last row, we show the reduced $\chi^{2}$ values considering all three parameters. Blank regions indicate parameter combinations that produce nonphysical outputs in the \citetalias{Fielding22} model, so they are ignored in the figures. Overall, we find that \etamcool\ is the most tightly constrained parameter, while \Mcloud\ and \etamhot\ are loosely constrained. See more discussion in Section \ref{sec:single}.}}
\label{fig:grid_model}
\end{figure*}

Next, we identify the key free parameters varied in the model. 
\cite{Fielding22} have identified the primary ones that govern the outputs. 
On the one hand, these include galaxy-scale properties, e.g., SFR, the depth of the gravitational potential (quantified by 
\Mstar\ or \Vcir), and the size of the star-forming region (\Rstar\ $\sim$ \Rhalf\ measured in UV). For our sample, these quantitatives are measured directly from observations \citep{Berg22, Xu22a}; therefore we fix them to the observed values (Table~\ref{tab:obs}). On the other hand, there are parameters that mark the initial status of the wind/outflows (i.e., their values at \Rstar). These include the initial hot-phase mass-loading factor (\etamhotinit $\coloneqq$ \Mdotwindinit/SFR), the initial cool-phase mass-loading factor (\etamcoolinit $\coloneqq$ \Mdotcloudinit/SFR), and the initial outflow cloud mass (\Mcloudinit). We select them because previous analyses have shown that UV absorption lines place tight constraints on the characteristic \etamcool\ and \Mcloud\ \citep{Xu22a, Xu23a}. Additionally, although the hot phase is not directly detected, its interaction with the cool phase may allow us to set limits on \etamhotinit. 

Overall, we fit the observed \Vcloud, \HWHMcloud, and \Nhcloud\ for each galaxy by varying three free parameters in the \citetalias{Fielding22} model, specifically with log(\etamhotinit) $\in$ ($-1.5$, $1.5$), log(\etamcoolinit) $\in$ ($-1.5$, $1.5$), and log[\Mcloudinit\ (\Msun)] $\in$ (1, 6.5). We also limit \Mcloudinit\ $<$ 10\% \Mstar\ of the galaxy.  The observed galaxy and outflow properties are listed in the first seven columns of Table~\ref{tab:obs}.

We perform the fits using a Bayesian framework implemented via the \texttt{emcee} Markov Chain Monte Carlo (MCMC) sampler \citep{Foreman-Mackey13}.  We sample the posterior probability distribution using $N = 72$ walkers for a total chain length of $20\times\tau_{\mathrm{max}}$, where $\tau_{\mathrm{max}}$ is the largest integrated autocorrelation time across all parameters. This is to ensure adequate convergence and independent sampling. The resulting posterior distributions provide both median estimates and 16th/84th percentile credible intervals for all parameters. The fitting results are listed in the last four columns of Table~\ref{tab:obs} and discussed in Sections~\ref{sec:single} and \ref{sec:whole}.

\subsection{Scope of the Paper}

While the \citetalias{Fielding22} model represents a significant step forward in developing simple and physically motivated multiphase wind models, it necessarily assumes time independence and a specific form for the cloud–wind interaction, both of which are approximations to the full problem. In addition, several parameters in that model are held fixed at fiducial values because the present data do not have the leverage to constrain the full parameter space. These include the normalization of the turbulent mixing–driven cloud mass growth rate ($f_{\dot{M}}=0.33$) and the parameter setting the strength of turbulence in the TRML relative to the cloud–wind relative velocity ($f_{\rm turb}=0.1$). Therefore, the results presented in the following sections should be interpreted as effective, model-conditional constraints. Future papers will relax some of these assumptions and explore a broader and more flexible model hierarchy to assess how robust these inferences are to changes in the underlying physics.

\section{Results for an Individual Galaxy}
\label{sec:single}

In this Section, we choose a representative galaxy, J0021+0052, to help explain the fitting results. It has SFR = 11.75 \Msun\ yr$^{-1}$, \Mstar\ = 10$^{9}$ \Msun, and \Rhalf\ = 0.45 kpc \citep{Xu22a}. In Figure~\ref{fig:single_model}, we show its fitting results. The left panels display the UV absorption lines used to derive the velocity–resolved \dNhdv\ profile in the right panel (blue). The 1$\sigma$ uncertainties are indicated by shaded regions. The best-fit \dNhdv\ curve (in red), computed with the settings in Section~\ref{sec:param}, matches the data within the 1$\sigma$ uncertainties. To further interpret the fitting results, we address two key questions as follows.

\subsection{How well are the Hot and Cool Phase Outflow Parameters Constrained?}
\label{sec:Q1}

First, we assess how well the three free parameters (\etamhotinit, \etamcoolinit, and \Mcloudinit) are constrained by the data. We explore it with a three-dimensional grid of models. Figure~\ref{fig:grid_model} visualizes the parameter space: each panel scans two parameters on the axes, while the third is fixed to the galaxy’s best-fit value. 
The first three rows show model predictions for \Vcloud, \HWHMcloud, and \Nhcloud\ as the color scales, indicating the constraining power of each observable. The bottom row shows the reduced \chisq\ surfaces. Black stars mark the best-fit model location for J0021+0052. Blank regions denote parameter combinations that produce nonphysical solutions in \citetalias{Fielding22} and are excluded from further consideration.

We note that \etamcoolinit\ is tightly constrained, with uncertainties $<$ 0.1 dex. This is most evident in panel 10 of Figure \ref{fig:grid_model}. Such tight constraints are expected because the fitted moments of \dNhdv\ can be mapped directly to \Mdotcloud\ (Equation \ref{eq:Mdot}) and, for a fixed SFR, to \etamcool. By contrast, \Mcloudinit\ is less tightly—but still moderately—constrained, with typical uncertainties of 0.2 -- 0.3 dex (e.g., panels 3, 6, and 12). This weaker constraint arises because \Mcloud\ scales with $\Nh$ but also depends on additional factors, including the outflow hydrogen number density ($\nh$) and and how outflow clouds shield each other in the line-of-sight (LOS) \citep[see Section 4.4 in][]{Xu23a}.

\etamhotinit\ is moderately constrained with uncertainties of $\sim0.3$ dex (e.g., panels 1, 4, 10 of Figure~\ref{fig:grid_model}). Among the three observables, \Vcloud\ carries more leverage when \etamhotinit\ $>1$, whereas \HWHMcloud\ is more informative when \etamhotinit\ $<1$. These can be seen from the changes of color gradients in the horizontal directions in panels 1 and 4, respectively. Overall, this supports the expectation that, although the observations probe only the cool outflow clouds, the hot wind can be constrained indirectly because the two phases interact as they propagate out of the galaxy. We reach a similar conclusion for other galaxies in our sample, which we discuss more in Section \ref{sec:whole}.

We note that $\Omega = 4\pi$ is assumed in our models, which is a statistical conclusion from \cite{Xu22a} for local starburst galaxies. Since \Mdot\ and the corresponding \etam\ are linearly proportional to $\Omega$ (see Equation \ref{eq:Mdot}), our derived \etamcoolinit\ and \etamhotinit\ can be viewed as angle averaged quantities. For example, if a galaxy actually have $\Omega = 2\pi$, then both \etam\ should drop by a factor of two. We also find the model fit with a smaller $\Omega$ yield larger \Mdotcloudinit. This is likely because reducing $\Omega$ decreases the total number of clouds intersecting the line of sight, which can be compensated by larger cloud masses to reproduce the observed column density distribution.

\begin{figure*}
\center
	\includegraphics[angle=0,trim={0.0cm 0.0cm 0cm 0.0cm},clip=true,width=1.0\linewidth,keepaspectratio]{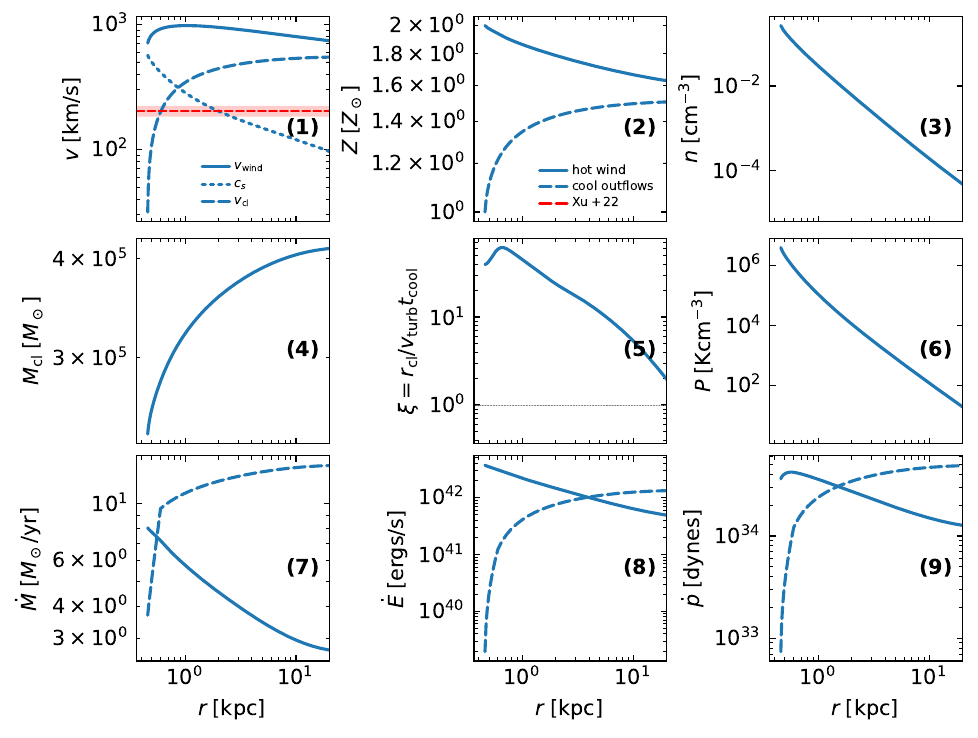}

\caption{\normalfont{Radial Distributions of galactic wind and outflow properties from the best-fit \citetalias{Fielding22} model for J0021+0052. In each panel, we show the hot wind properties in solid blue lines and outflow cloud properties in long dashed blue lines. For left-top to right-bottom, we show the radial distributions for \textbf{(1)} gas velocity, \textbf{(2)} gas metallicity, \textbf{(3)} gas density, \textbf{(4)} mass of outflow clouds, \textbf{(5)} ratio of mixing time to cooling time for outflow clouds, \textbf{(6)} gas pressure, \textbf{(7)} mass outflow rates, \textbf{(8)} energy outflow rates, and \textbf{(9)} momentum outflow rates. We also overlay the observed outflow velocity for this galaxy \citep[using spatially-integrated spectra,][]{Xu22a} as red dashed lines with red shades represent the 1$\sigma$ error ranges. }}
\label{fig:radial_dist}
\end{figure*}

\subsection{How to Resolve the Outflow Features using Spatially-integrated Measurements?}
\label{sec:Q2}

Having established how parameters are constrained, we now ask whether spatially integrated spectra can recover spatial information about the outflow. This is plausible because \dNhdv\ derived from absorption lines should already encode spatial information about the outflow structure \citep[e.g.,][]{Carr18, Carr25, Huberty24, Peng24, Xu25}. Specifically, clouds at smaller radii are generally denser and move more slowly, contributing absorption at lower velocities. Conversely, clouds at larger radii are more diffuse and travel faster, producing absorption at higher velocities. High spectral resolution in absorption-line profiles therefore helps constrain the radial distribution of the outflowing gas. We illustrate this mapping (from velocity to radius) by annotating the model-predicted radius along the top-axis of the right panel of Figure~\ref{fig:single_model}. 

We note that the velocity–radius mapping is strictly one-to-one only when the bulk velocity field is monotonic. In more general cases with non-monotonic flows, the inversion from velocity to radius is no longer unique, as gas at multiple radii can contribute to the same observed velocity. Nevertheless, the absorption profile still encodes useful information about the global distribution of the outflowing gas (e.g., total column density and characteristic velocity scales), although their association with specific radii becomes model-dependent. For the galaxies modeled in this paper, all systems are well-fit with monotonic velocity distributions (Section~\ref{sec:sample_spatial_dist}), so the mapping remains well-defined. We also emphasize that the \citetalias{Fielding22} framework does not require monotonic velocity profiles; monotonicity simply emerges from the best-fit solutions for this sample.

To clarify the interaction between the hot and cool phases and their radial structure, we present radial profiles of key quantities in Figure~\ref{fig:radial_dist}. The blue curves come directly from the \citetalias{Fielding22} model using the best-fit solution for J0021+0052, where solid and dashed lines denote the hot wind and the cool outflow clouds, respectively. Panel (1) shows a rapid acceleration of the cool outflow from \Rhalf\ (0.45 kpc for J0021+0052) to $\sim$ 1 kpc, followed by a plateau at larger radii. The hot-wind velocity, \Vwind, also exhibits a slight deceleration beyond 1~kpc. These trends are consistent with continued mass entrainment from the hot phase by the cool clouds, leading to an increase in the cloud mass (panel 4). Notably, the hot wind remains supersonic at all radii. This behavior matches the criterion that \citet{Nguyen24} suggest for 1D time-steady analytic wind solutions to reproduce 3D time-dependent calculations. In red, we overlay the observed outflow velocity, \Vcloud\ (see Section~\ref{sec:param}), measured from the spatially integrated HST/COS spectrum \citep{Xu22a}, with shaded bands indicating the 1$\sigma$ uncertainties. We find it intersects the modeled \Vcloud(r) at $\sim$~0.7 -- 0.8~kpc, close to 2\Rhalf\ for this galaxy. 

For the whole sample, we perform similar comparisons (detailed in Section~\ref{sec:whole}) and find that the intersections consistently occur between 1 and 2\Rhalf. This suggests that the spatially integrated velocities derived from our observations correspond to the resolved velocities at a characteristic radius. This range is consistent with previous work that found outflow sizes extending to 1 -- 2\Rhalf, including \cite{Newman12b}, who directly measured outflow sizes from optical emission lines, and \cite{Xu23a} who indirectly derived outflow sizes using UV absorption lines.

The bottom row shows the radial profiles of the mass, momentum, and energy outflow rates. A clear transfer of mass, momentum, and energy from the hot phase (solid blue) to the cool phase (dashed blue) can be seen. Taken together, interaction between the cool and hot phases significantly shapes the properties of both.
The velocity–radius mapping also enables us to recover spatial information from integrated spectra.
In the next section, we extend this analysis to the full sample.

\section{Results for the Sample}
\label{sec:whole}

In the parent sample  described in Section \ref{sec:CLASSY}, we select 22 galaxies to conduct the fitting. We exclude ones that have (1) no UV outflow signatures or  are without a securely measured \dNhdv\ profile, and (2) galaxies with too low outflow velocities  such that the line profiles are not well-resolved by HST/COS. We find the latter mainly happens when the galaxy's \Rhalf\ is close to or larger than the COS aperture. Therefore, we likely only observe their cluster-scale outflows instead of the galactic-scale one. Thus, our model does not apply to them.

\begin{figure*}
\center
	\includegraphics[angle=0,trim={0.0cm 0.0cm 0cm 0.0cm},clip=true,width=1.0\linewidth,keepaspectratio]{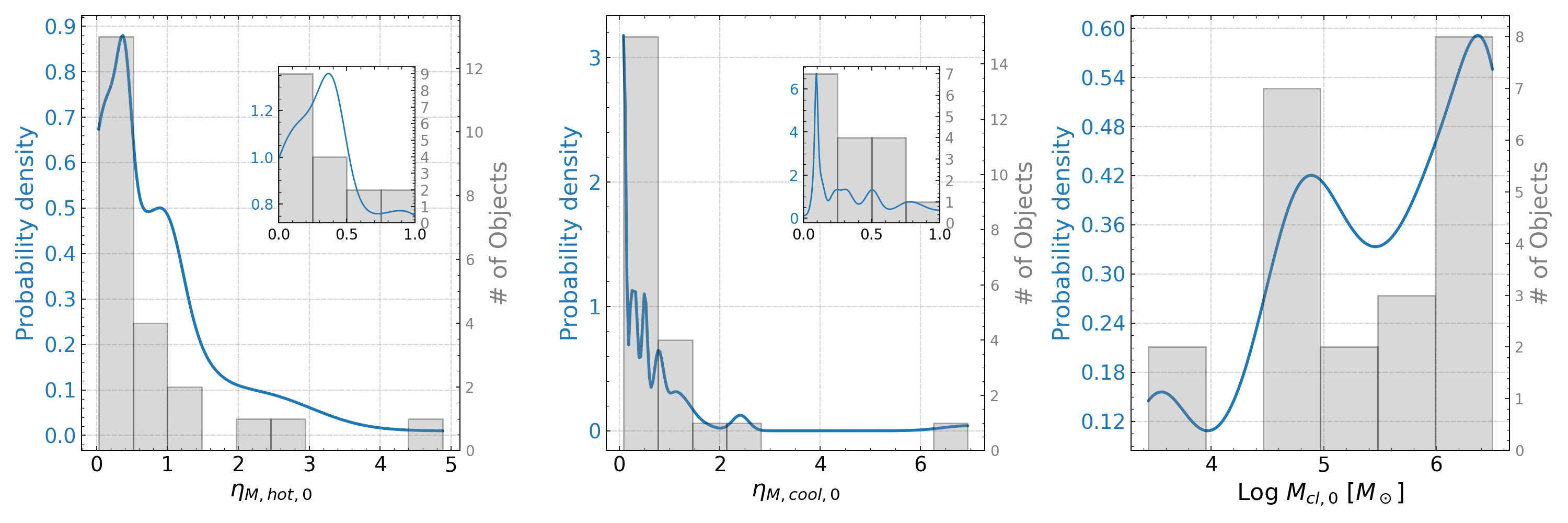}

\caption{\normalfont{Distributions of the best-fit parameters for galaxies in our sample. In blue, we show the probability density function (PDF) for each of the three parameters fit by the model. In gray, we show the distributions of the number of objects for the three parameters. The insets show the zoom-in regions between 0 -- 1 with finer binning on the x-axis.}}
\label{fig:distribution}
\end{figure*}

\subsection{Distributions of the Model Parameters}
\label{sec:sample_params_dist}

For all 22 galaxies in our sample, their observed galaxy and outflow measurements are listed in Table~\ref{tab:obs}, where the best-fit parameters (\etamhotinit, \etamcoolinit, \Mcloudinit) and reduced \chisq\ values are summarized in the last four columns.

Overall, we find most galaxies fit well with the models (\chisq $<$ 1.5 for 18 out of 22 galaxies), while the remaining four have relatively worse fit, which we discuss in Section \ref{sec:caveats}. We show the distributions of the fitting parameters in Figure \ref{fig:distribution} as histograms. We also overlay posterior probability density functions (PDF) from the MCMC to reflect parameter uncertainties. In the left panel, we find $\sim$ 60\% of our galaxies prefer a hot phase initial mass loading factors (\etamhotinit) $<$ 1.0. This is consistent with previous theoretical studies of \etamhot, e.g., in \cite{Zhang14}. They highlight that heavily mass-loaded winds cannot be described by the adiabatic CC85 model because they become strongly radiative. They suggest \etamhot\ $\lesssim$ 1 when SFR $\gtrsim$ 10 \Msun/yr, which is mostly true for our galaxies matching this criteria. In addition, there are three galaxies with large \etamhotinit\ $\gtrsim$ 10 (e.g., J0942+3547, J1150+1501). This is likely due to their very low star formation rates (SFRs; $<$ 0.2~\Msun~yr$^{-1}$) derived from SED fitting \citep{Berg22}, which can have larger relative uncertainties. Though this assumption was found to provide satisfying fits to the broadband photometry of our galaxies, more complex SFH (e.g., bursty star formation) may explain why we observe outflow signatures in these galaxies with very low SFR.

For \etamhot, \cite{Pandya21} found the following relationship for galaxies from the FIRE-2 simulations based on spatially integrated measurements:

\begin{equation}
\label{eq:etamhot}
\eta_{M,\mathrm{hot}} \;=\; 10^{2.4}\,\left(\frac{M_*}{M_\odot}\right)^{-0.27}
\end{equation}

This suggests that \etamhot\ $\lesssim$ 3 for the mass range of our galaxies (i.e., \Mstar\ $\gtrsim$ 10$^{7}$ \Msun), which is also consistent with what we find in Table \ref{tab:obs}. 

In the middle panel of Figure \ref{fig:distribution}, we show the cool phase initial mass loading factors (\etamcoolinit). We find most of our galaxies ($\sim$ 80\%) prefer \etamcoolinit\ $<$ 1, with a PDF that peaks $\sim$ 0.1. This is expected since the cool outflow clouds initially are at relatively low velocity and mass when they are not accelerated by the hot wind. 

In the right panel, we show that the initial mass of outflow clouds (\Mcloudinit) favors high values approaching $10^{6.5}$~\Msun. We caution that the current model assumes that all outflow clouds have the same mass, which is likely unrealistic. Therefore, we leave the interpretation of \Mcloud\ to the subsequent paper, when we will update the model with more physically motivated mass distributions of cool outflow clouds.

\begin{figure*}
\center
	\includegraphics[angle=0,trim={0.0cm 0.0cm 0cm 0.0cm},clip=true,width=0.5\linewidth,keepaspectratio]{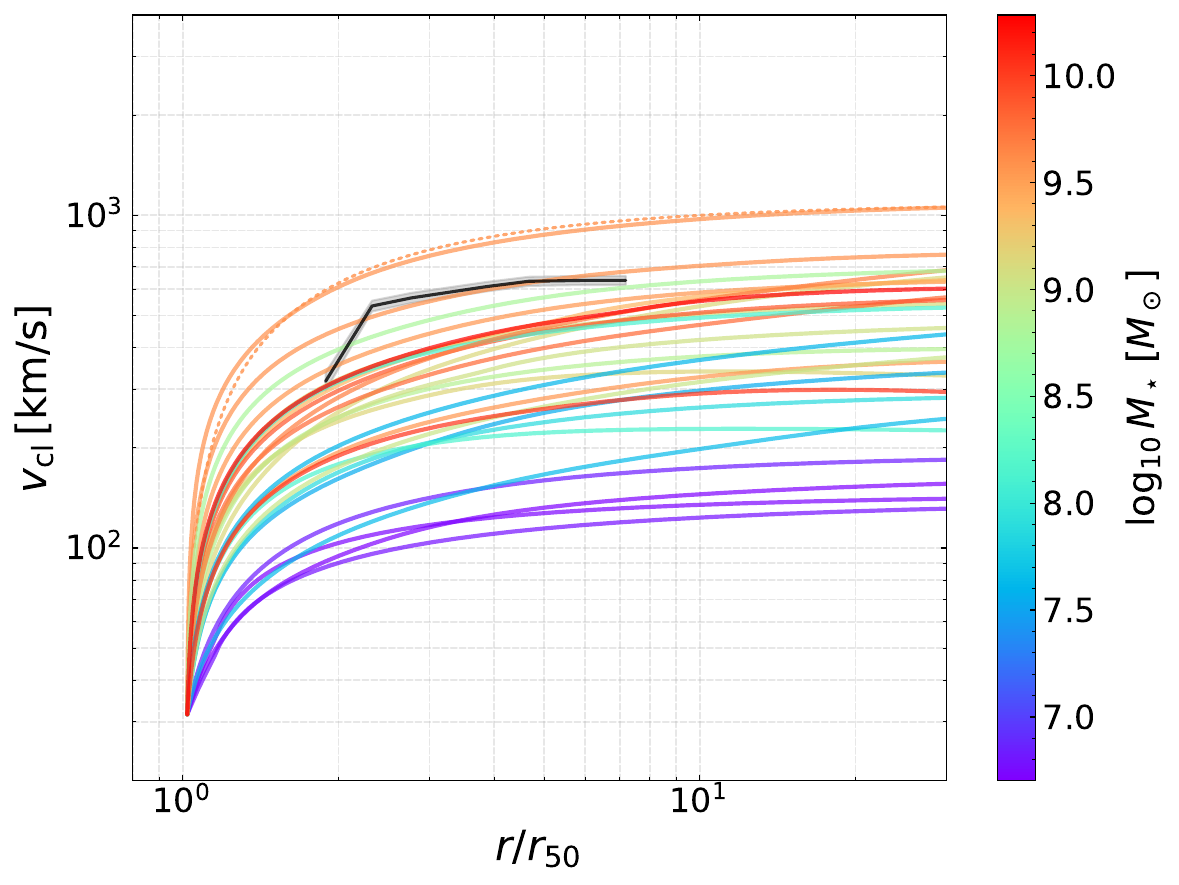}
	\includegraphics[angle=0,trim={0.0cm 0.0cm 0cm 0.0cm},clip=true,width=0.5\linewidth,keepaspectratio]{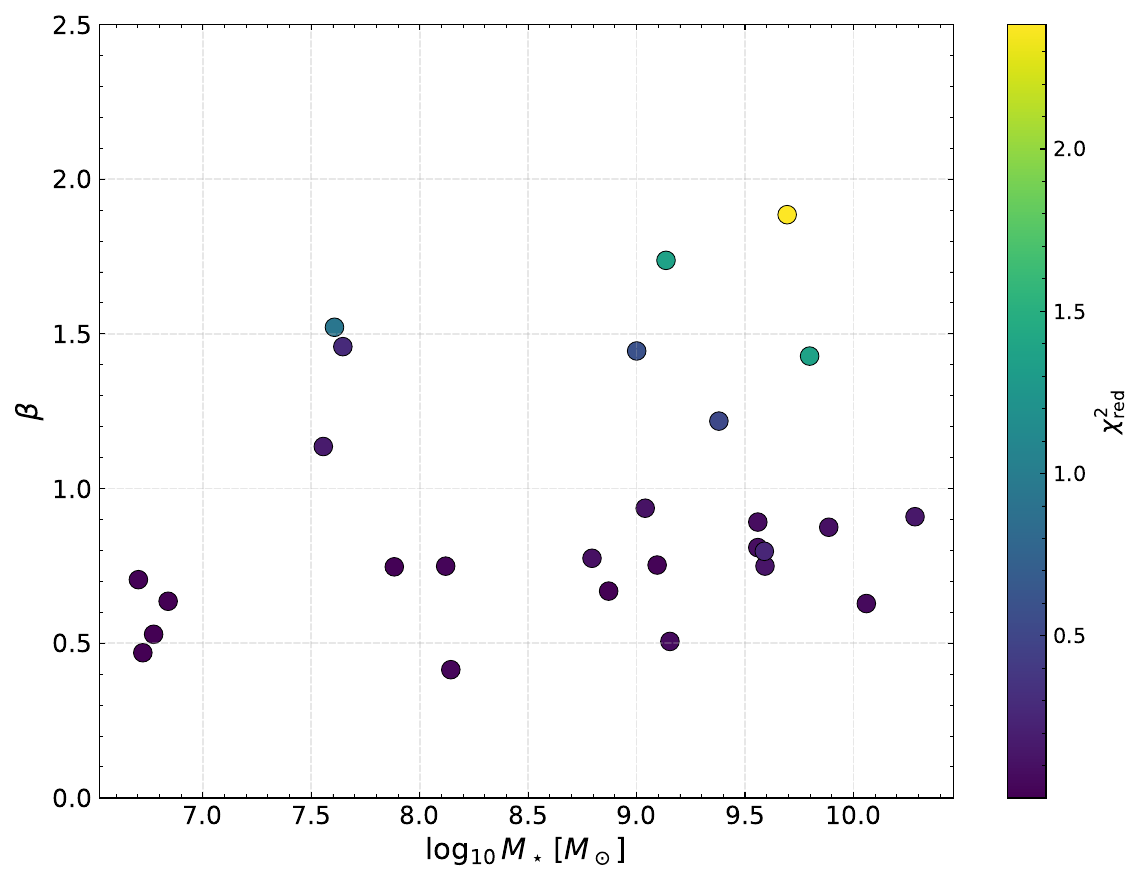}
\caption{\normalfont{\textbf{Left:} Radial distribution of cool outflows. Each colored line shows the best-fit \protect{\citetalias{Fielding22}} model for an individual galaxy. The x-axis is normalized by the galaxy’s half-light radius (\rhalf), and the lines are color-coded by stellar mass (\Mstar). Overlaid in black is the spatially resolved velocity profile of cool outflowing clouds from the local starburst galaxy M~82, with 1$\sigma$ uncertainties shown as gray-shaded regions \citep{Xu23c}. An illustrative example of the $\beta$-law fit (Equation~\ref{eq:v_r}) is shown for J1416+1223 (orange dashed line), where $v_\infty = 1087$~\kms, \Rhalf\ = 0.27~kpc, and $\beta = 0.8$. It closely follows the corresponding \protect{\citetalias{Fielding22}} profile (top solid orange line). \textbf{Right:} Best-fit $\beta$ values for the full sample, most of which fall in the range $\beta = 0.5$ -- 1.5, consistent with the M~82 profile ($\beta = 1$). See Section~\ref{sec:sample_spatial_dist} for discussion.}}
\label{fig:spatial_dist_vel}
\end{figure*}

\begin{figure*}
\center
	\includegraphics[angle=0,trim={0.0cm 0.0cm 0cm 0.0cm},clip=true,width=0.5\linewidth,keepaspectratio]{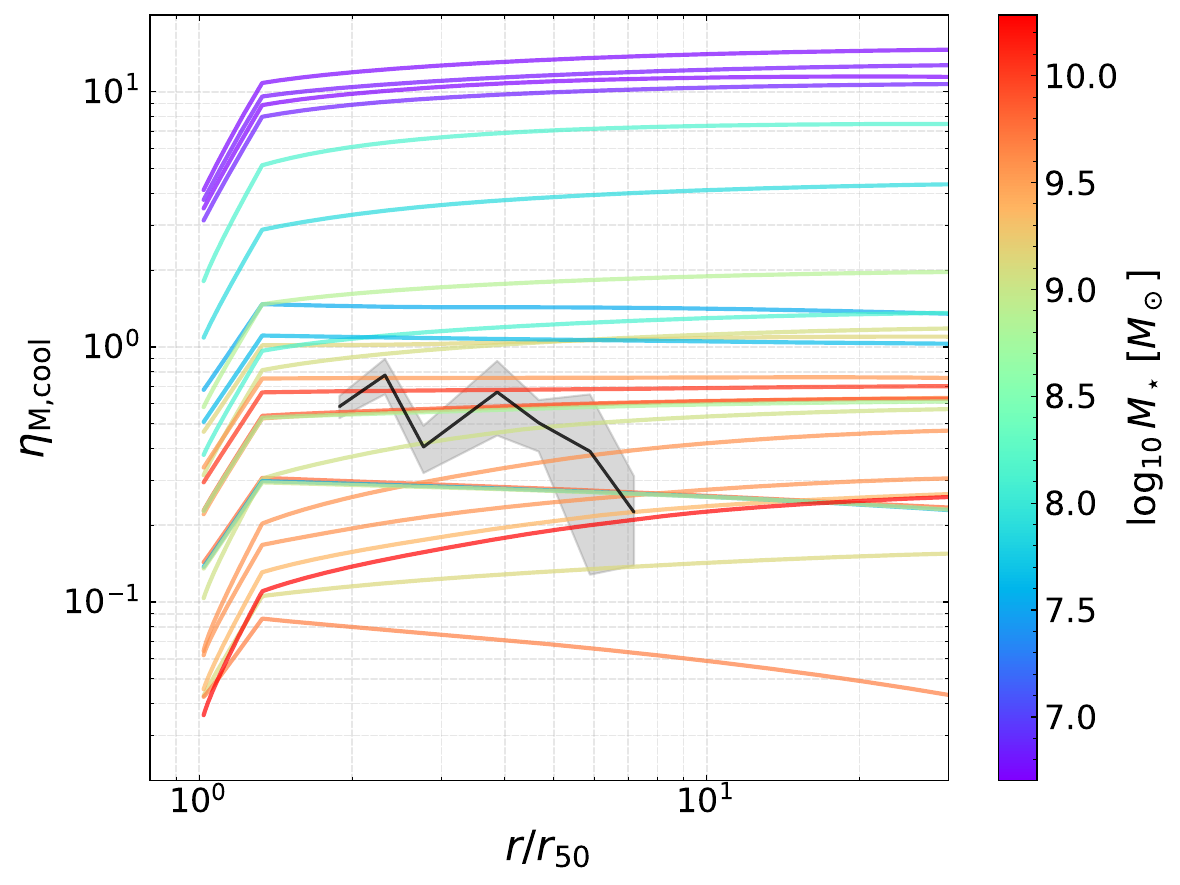}
	\includegraphics[angle=0,trim={0.0cm 0.0cm 0cm 0.0cm},clip=true,width=0.5\linewidth,keepaspectratio]{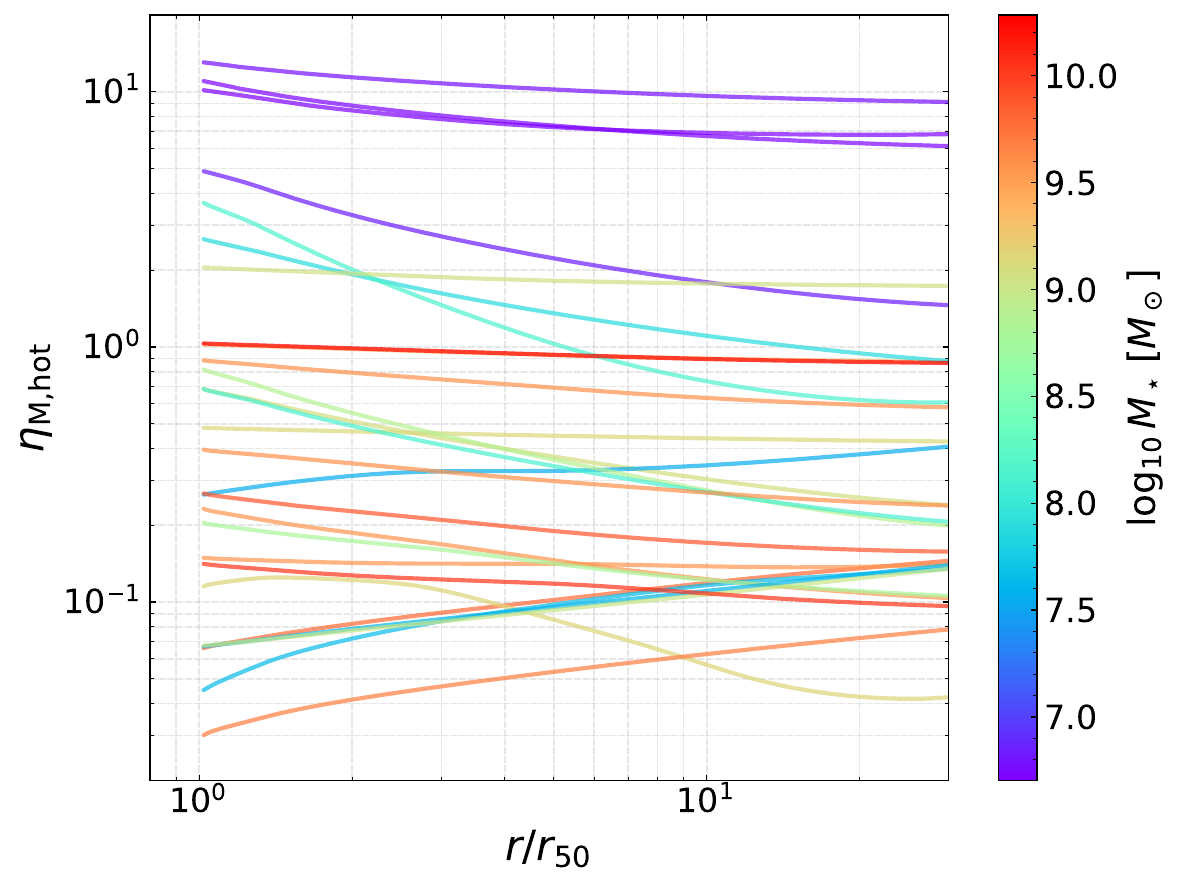}
\caption{\normalfont{Same as Figure \ref{fig:spatial_dist_vel} but for mass loading factors' variations over radius for cool outflow clouds (\textbf{left}) and hot wind (\textbf{right}) derived from the model.}}
\label{fig:spatial_dist_Xdot}
\end{figure*}

\begin{figure*}
\center
	\includegraphics[angle=0,trim={0.0cm 0.0cm 0cm 0.0cm},clip=true,width=0.5\linewidth,keepaspectratio]{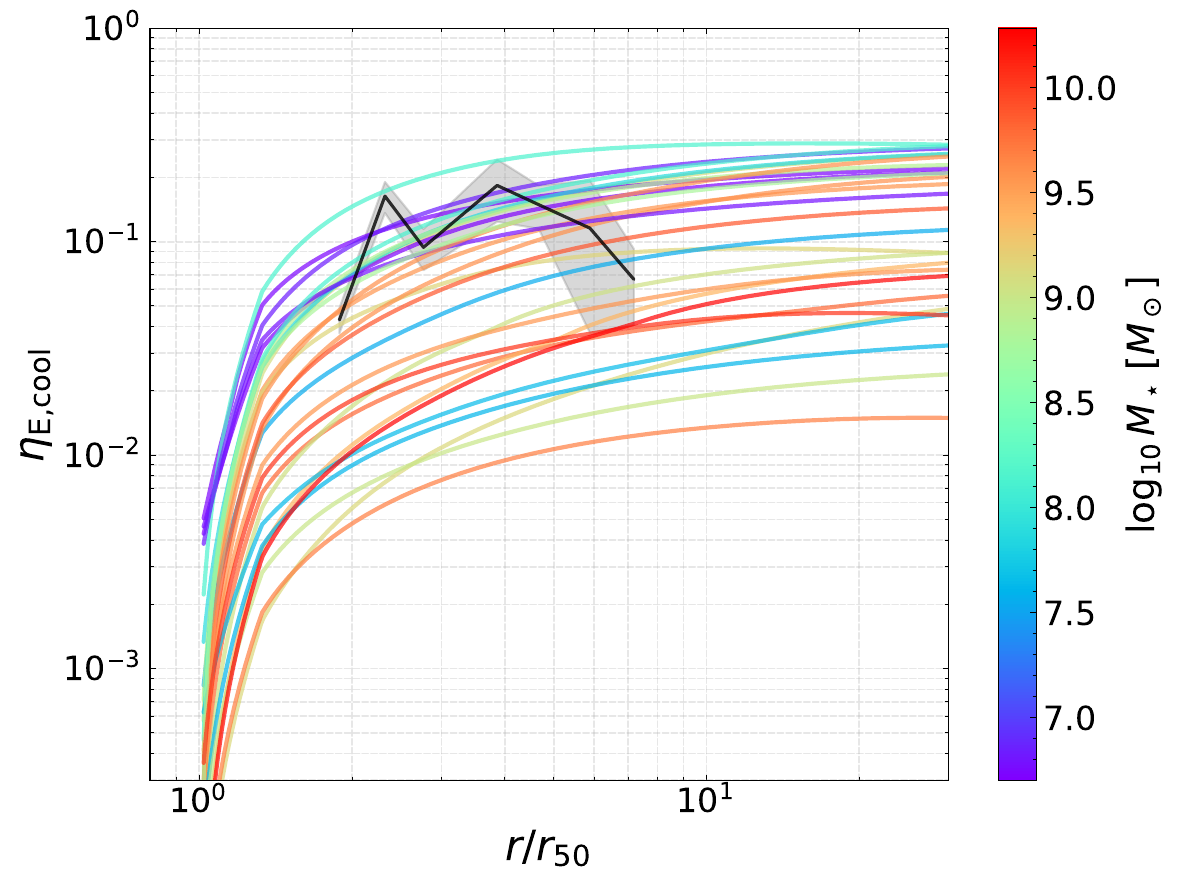}
	\includegraphics[angle=0,trim={0.0cm 0.0cm 0cm 0.0cm},clip=true,width=0.5\linewidth,keepaspectratio]{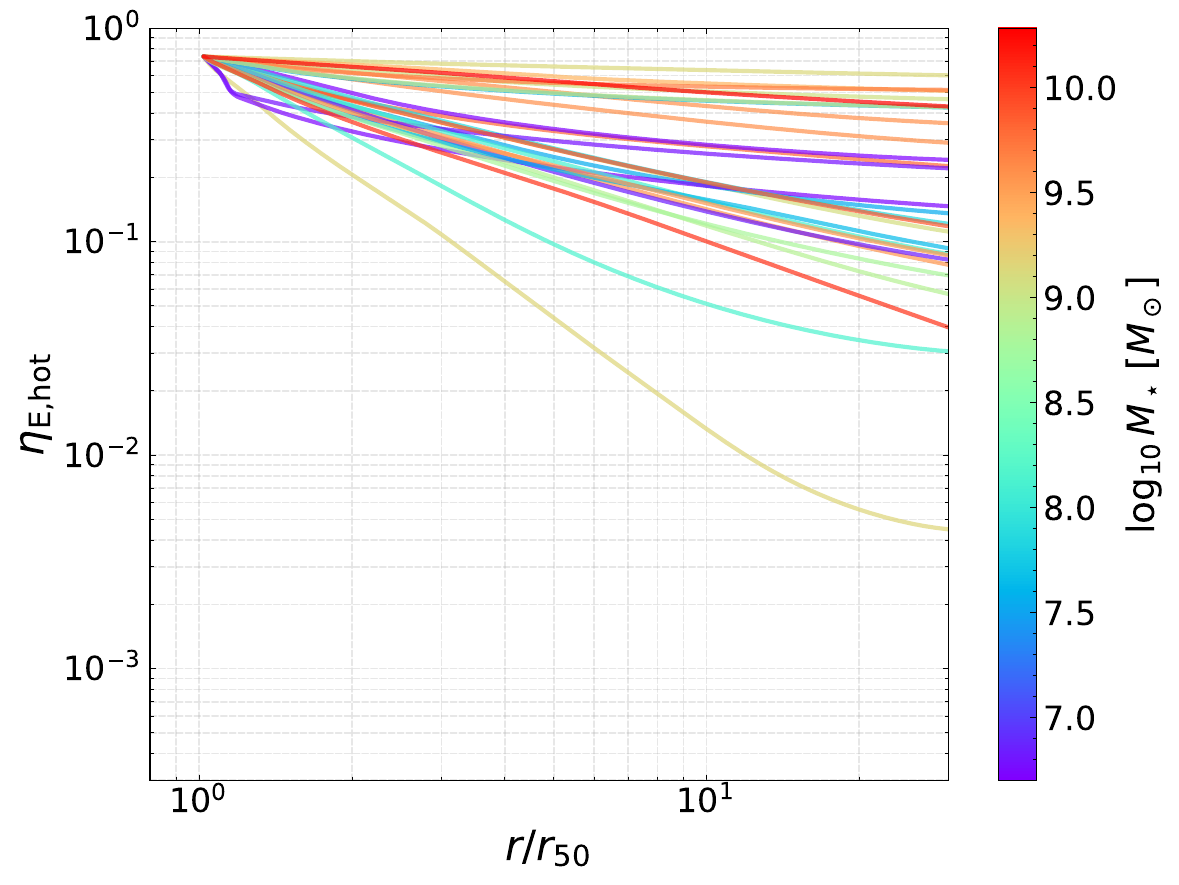}
    
    \makebox[\linewidth][l]{%
      \includegraphics[trim={0 0 0 0},clip,width=0.5\linewidth]{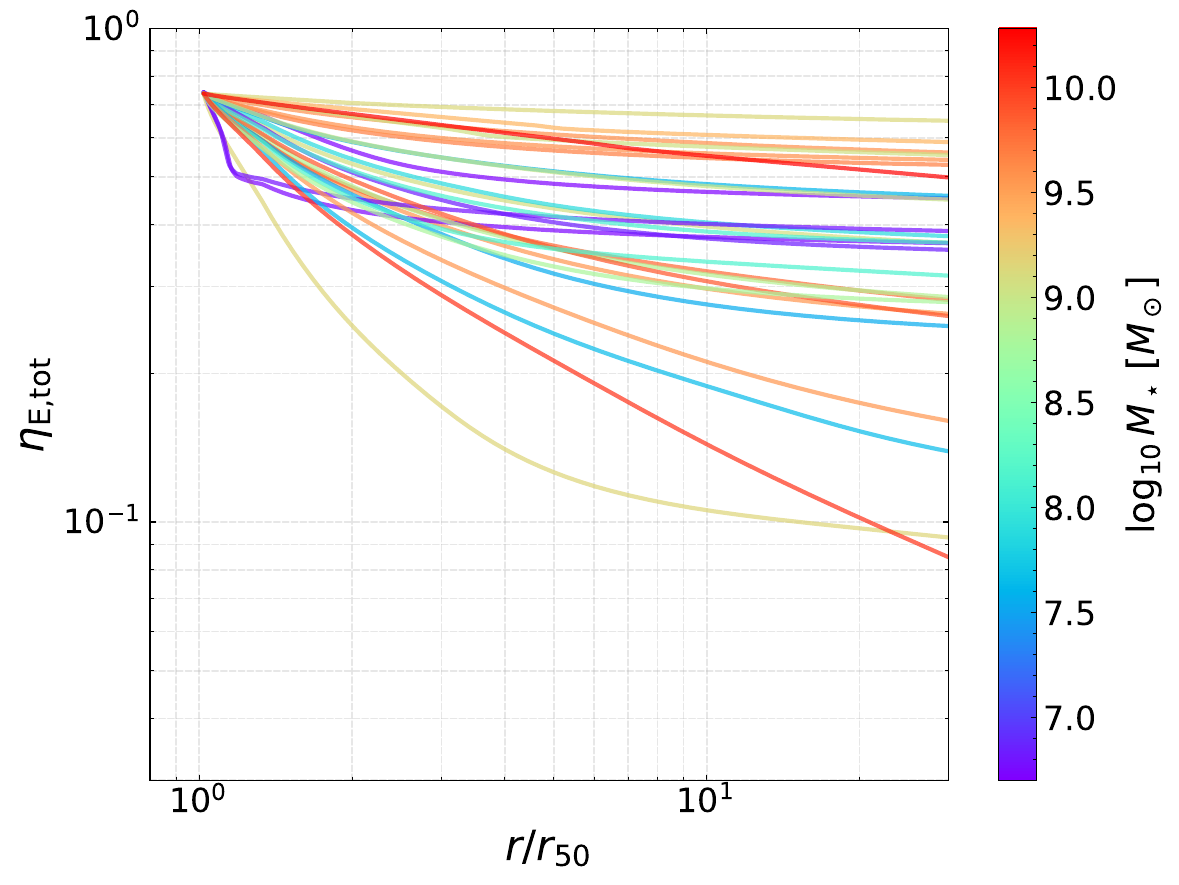}%
    }
\caption{\normalfont{Same as Figure \ref{fig:spatial_dist_vel} but for energy loading factors for the cool outflow clouds (\textbf{top-left}), hot wind (\textbf{top-right}), and the sum of both (\textbf{bottom-left}).}}
\label{fig:spatial_dist_Etot}
\end{figure*}

\subsection{Radial Variations of Outflow Properties}
\label{sec:sample_spatial_dist}

As in Section~\ref{sec:Q2}, we present our best-fit-model predicted radial distributions of outflow properties for all galaxies in Figures~\ref{fig:spatial_dist_vel} -- \ref{fig:spatial_dist_Etot}. Each curve corresponds to one galaxy and is color-coded by \Mstar\ of that galaxy. The horizontal axis shows radius normalized by its \Rhalf.

In Figure~\ref{fig:spatial_dist_vel}, one empirical finding is that the cool-phase outflow velocity, \Vcloud, increases monotonically with radius for all galaxies (solid lines). This behavior is consistent with the cool outflow's acceleration driven by the hot wind through mainly momentum transfer. The non-detection of decelerating outflows is likely because such outflowing clouds lose mass and persist for shorter time scales, making them harder to be observed. There is a strong trend in which more massive galaxies show higher \Vcloud, which is known through the outflow scaling relationships in previous studies \citep[e.g.,][]{Heckman15, Xu22a}. 

We also overlay the observed spatially resolved velocity profile from M~82 as a black line with its 1$\sigma$ error in gray shaded regions.  This profile approximates a $\beta$-law \citep{Xu23c}:

\begin{equation}\label{eq:v_r}
    \begin{aligned}
    v(r)    &= v_\infty \left (1 - r_\text{50}/r  \right )^{\beta}
    \end{aligned}
\end{equation}
where v$_\infty$ is the terminal outflow velocity and $\beta$ is the power index. We find that the M~82 profile exhibits a similar trend to those in our galaxies—characterized by a rapid acceleration phase transitioning to an asymptotic velocity plateau.

To further assess this similarity, we fit each curve by varying $v_\infty$ and $\beta$ while fixing \Rhalf\ to the observed values. An illustrative example is shown for J1416+1223 (orange dashed line), where a $\beta$-law with $v_\infty = 1087$~\kms, \Rhalf\ = 0.27~kpc, and $\beta = 0.8$ closely traces the inferred model curve (the top orange solid line). The right panel of Figure~\ref{fig:spatial_dist_vel} presents the results for the full sample. We find that nearly all \Vcloud($r$) profiles are well-reproduced using $\beta = 0.5$ -- 1.5, consistent with the resolved outflow structure observed in M~82. These are different from the alternative forms of  \Vcloud($r$) proposed in other studies, e.g., \cite{Huberty24} adopts $v(r) \propto (r/r_\text{50})^{\gamma}$. We highlight the need for future large samples of spatially resolved observations to definitively constrain the underlying outflow velocity structure.

In Figure~\ref{fig:spatial_dist_Xdot}, we present the radial distributions of \etam$(r)$, the outflow mass loading factors for the cool and hot phases, separately. They are calculated from the radial dependent mass loading factor (Figure \ref{fig:radial_dist}) divided by the SFR of the galaxy. The majority of the galaxies show a monotonically increasing \etamcool$(r)$ and a decreasing \etamhot$(r)$ with radius. There are a few exceptions that show completely the reverse trend (J1414+0540, J1448--0110, J1521+0759 in both panels, and J0150+1308, J0942+3547 in the right panel). We do not find special properties of them versus other galaxies, except they have relatively high fitted values for \Mcloudinit\ ($\gtrsim10^5$ \Msun), but we note our adopted single value for \Mcloudinit\ is likely not realistic (see discussion in Section \ref{sec:caveats}). It is likely that those outflows fail to be accelerated due to their low mass loading. This pattern has also been predicted in previous simulations \citep[e.g.,][]{Schneider20}, which we discuss in Section \ref{sec:discuss}. In both panels, we also see that more massive galaxies tend to have smaller loading factors, which is consistent with known scaling relationships of outflows \citep[e.g.,][]{Heckman15}.

\begin{figure}
\center
	\includegraphics[angle=0,trim={0.0cm 0.0cm 0cm 0.0cm},clip=true,width=1.0\linewidth,keepaspectratio]{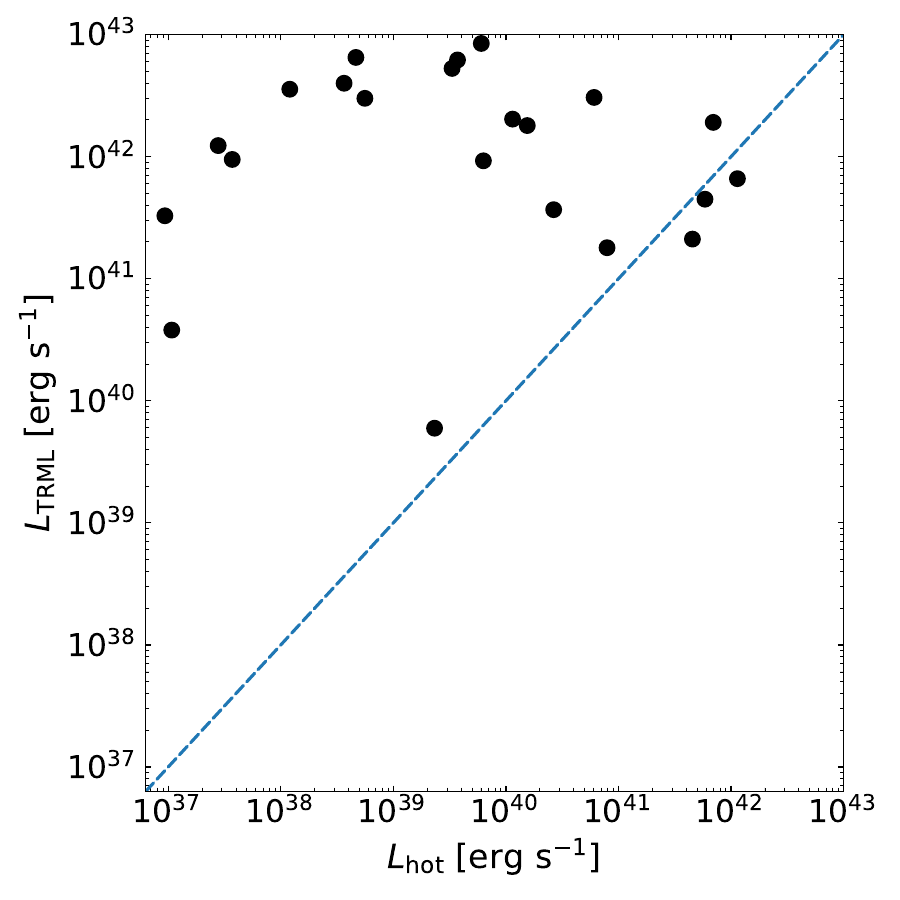}

\caption{\normalfont{
Radiative energy losses for all of our galaxies using the best-fit model. The radiative luminosity from turbulent radiative mixing layers ($L_{\rm TRML}$) is plotted against the radiative cooling luminosity of the hot phase ($L_{\rm hot}$). The dashed line shows the one-to-one relation. Most galaxies lie above this line, indicating that cooling associated with TRML dominates.
}}
\label{fig:energy_loss}
\end{figure}

In Figure~\ref{fig:spatial_dist_Etot}, we present the variation of energy loading factor over radius (\etaE$(r)$), which is the energy-outflow rates normalized by the star-formation energy injection rate (\EdotStar). \EdotStar\ is computed using Starburst99 models \citep{Leitherer99}, yielding \EdotStar\ = 4.3~$\times 10^{41}$~SFR~ergs s$^{-1}$. In the top two panels, we show \etaE$(r)$\ for the cool and hot phases, separately. In all galaxies, we find \etaecool$(r)$\ (or \etaehot$(r)$) increases (or decreases) with radius. This is again expected given energy transfer from the hot to cool phases. In the bottom panel, we sum the energy loading factors from the two phases to check the total energy of the system. We find the multiphase wind+outflow component lose substantial amounts (10 -- 90\%) of their initial energy while they move out of the galaxies. In Figure \ref{fig:energy_loss}, we compare the energy loss in the two most likely scenarios, i.e., in TRML ($L_{\rm TRML}$) or through radiative cooling in the volume-filling hot phase ($L_{\rm hot}$). It is clear that the former dominates. This result suggests that turbulent mixing between the hot wind and entrained cool clouds provides the primary channel for radiative energy loss in multiphase galactic winds.

Overall, the interactions between the cool and hot phases in galactic winds significantly affect the radial distributions of both components. While most of our results are consistent with empirical expectations, some rare cases show unanticipated results. We do not pursue these cases further to avoid over-interpretation and we will return to these questions in Paper II. In addition, the substantial mass, momentum, and energy exchanges between the hot wind to the cool clouds also underscores that galactic-wind models and simulations need to capture both phases. Otherwise, inferences from single-phase treatments can be biased.

\begin{deluxetable}{lcccllrrrrc}
\tablecaption{Model Parameters and Results for the Sample \label{tab:obs}}
\tablehead{
\colhead{Object} & 
\colhead{SFR} & 
\colhead{\Rhalf} & 
\colhead{log(\Mstar)} & 
\colhead{\Vcloud} & 
\colhead{\HWHMcloud} & 
\colhead{log(\Nhcloud)} &
\colhead{log(\etamhotinit)} & 
\colhead{log(\etamcoolinit)} & 
\colhead{log(\Mcloudinit)} &
\colhead{$\chi^{2}$} \\
\colhead{} &
\colhead{(\Msun\ yr$^{-1}$)} & 
\colhead{(kpc)} & 
\colhead{(\Msun)} & 
\colhead{(km s$^{-1}$)} & 
\colhead{(km s$^{-1}$)} & 
\colhead{(cm$^{-2}$)} &
\colhead{} &
\colhead{} &
\colhead{(\Msun)} &
\colhead{}\\
\colhead{(1)} &
\colhead{(2)} &
\colhead{(3)} &
\colhead{(4)} &
\colhead{(5)} &
\colhead{(6)} &
\colhead{(7)} &
\colhead{(8)} &
\colhead{(9)} &
\colhead{(10)} &
\colhead{(11)}
}
\startdata
J0021+0052 & 11.75 & 0.45 & 9.09 & 366$^{+12}_{-181}$ & 158$^{+35}_{-21}$ & 20.71$^{+0.06}_{-0.05}$ & 0.36$^{+0.04}_{-0.36}$ & --0.29$^{+0.13}_{-0.08}$ & 6.27$^{+0.20}_{-0.59}$ &3.38 \\
J0036\textminus{}3333 & 10.23 & 0.12 & 9.14 & 150$^{+13}_{-11}$ & 83$^{+18}_{-13}$ & 20.45$^{+0.02}_{-0.02}$ & --0.30$^{+0.29}_{-0.46}$ & --1.00$^{+0.06}_{-0.05}$ & 6.09$^{+0.28}_{-0.45}$ &0.36 \\
J0055\textminus{}0021 & 42.66 & 0.57 & 9.56 & 466$^{+11}_{-253}$ & 219$^{+29}_{-48}$ & 20.92$^{+0.04}_{-0.04}$ & 0.18$^{+0.05}_{-0.44}$ & --0.48$^{+0.16}_{-0.06}$ & 6.23$^{+0.25}_{-0.36}$ &9.46 \\
J0150+1308 & 31.62 & 0.63 & 9.80 & 175$^{+12}_{-18}$ & 125$^{+17}_{-14}$ & 20.63$^{+0.04}_{-0.04}$ & --0.53$^{+0.33}_{-0.30}$ & --0.50$^{+0.05}_{-0.05}$ & 5.99$^{+0.36}_{-0.29}$ &0.80 \\
J0823+2806 & 30.20 & 0.25 & 9.38 & 199$^{+16}_{-10}$ & 164$^{+16}_{-10}$ & 20.44$^{+0.02}_{-0.02}$ & --0.20$^{+0.12}_{-0.17}$ & --0.89$^{+0.08}_{-0.06}$ & 6.02$^{+0.26}_{-0.37}$ &3.36 \\
J0938+5428 & 11.22 & 0.51 & 9.15 & 153$^{+10}_{-14}$ & 36$^{+41}_{-6}$ & 20.78$^{+0.06}_{-0.06}$ & --0.08$^{+0.27}_{-0.26}$ & --0.01$^{+0.06}_{-0.07}$ & 6.18$^{+0.23}_{-0.45}$ &0.27 \\
J0942+3547 & 0.17 & 0.10 & 7.56 & 100$^{+13}_{-11}$ & 58$^{+12}_{-11}$ & 20.03$^{+0.02}_{-0.02}$ & 0.44$^{+0.21}_{-0.40}$ & 0.12$^{+0.10}_{-0.08}$ & 4.62$^{+0.26}_{-0.34}$ &0.61 \\
J1024+0524 & 1.62 & 0.27 & 7.88 & 103$^{+4}_{-3}$ & 56$^{+14}_{-10}$ & 20.91$^{+0.02}_{-0.02}$ & 0.15$^{+0.17}_{-0.30}$ & 0.42$^{+0.03}_{-0.03}$ & 6.30$^{+0.14}_{-0.25}$ &0.27 \\
J1025+3622 & 10.96 & 0.79 & 8.87 & 149$^{+139}_{-8}$ & 101$^{+15}_{-20}$ & 20.79$^{+0.05}_{-0.04}$ & --0.20$^{+0.15}_{-0.27}$ & 0.11$^{+0.08}_{-0.04}$ & 6.29$^{+0.14}_{-0.48}$ &0.09 \\
J1112+5503 & 39.81 & 0.47 & 9.59 & 389$^{+22}_{-14}$ & 126$^{+60}_{-33}$ & 20.46$^{+0.03}_{-0.03}$ & --0.35$^{+0.35}_{-0.39}$ & --0.59$^{+0.13}_{-0.30}$ & 3.75$^{+0.61}_{-0.33}$ &0.69 \\
J1144+4012 & 32.36 & 0.91 & 9.89 & 177$^{+9}_{-8}$ & 118$^{+14}_{-15}$ & 20.64$^{+0.03}_{-0.03}$ & --0.40$^{+0.21}_{-0.36}$ & --0.32$^{+0.04}_{-0.04}$ & 6.13$^{+0.23}_{-0.35}$ &0.04 \\
J1150+1501 & 0.05 & 0.07 & 6.84 & 71$^{+6}_{-5}$ & 32$^{+10}_{-6}$ & 20.54$^{+0.03}_{-0.03}$ & 0.68$^{+0.22}_{-0.23}$ & 0.82$^{+0.03}_{-0.04}$ & 5.17$^{+0.34}_{-0.26}$ &0.75 \\
J1157+3220 & 9.33 & 0.65 & 9.04 & 161$^{+18}_{-12}$ & 47$^{+30}_{-8}$ & 20.13$^{+0.01}_{-0.01}$ & 0.24$^{+0.12}_{-0.27}$ & --0.60$^{+0.07}_{-0.05}$ & 6.04$^{+0.30}_{-0.41}$ &0.04 \\
J1200+1343 & 5.62 & 0.23 & 8.12 & 300$^{+19}_{-115}$ & 198$^{+10}_{-54}$ & 20.88$^{+0.06}_{-0.15}$ & 0.44$^{+0.01}_{-0.33}$ & --0.13$^{+0.05}_{-0.04}$ & 6.19$^{+0.24}_{-0.70}$ &2.16 \\
J1253\textminus{}0312 & 3.63 & 0.39 & 7.65 & 123$^{+12}_{-26}$ & 39$^{+21}_{-12}$ & 20.85$^{+0.03}_{-0.03}$ & --0.12$^{+0.46}_{-0.36}$ & 0.24$^{+0.05}_{-0.06}$ & 6.12$^{+0.28}_{-0.43}$ &1.19 \\
J1414+0540 & 5.00 & 0.36 & 9.69 & 178$^{+12}_{-9}$ & 107$^{+19}_{-19}$ & 20.39$^{+0.02}_{-0.01}$ & --0.56$^{+0.26}_{-0.32}$ & --1.02$^{+0.05}_{-0.04}$ & 6.26$^{+0.19}_{-0.25}$ &1.26 \\
J1416+1223 & 37.15 & 0.27 & 9.59 & 382$^{+20}_{-19}$ & 111$^{+6}_{-4}$ & 20.07$^{+0.02}_{-0.02}$ & --0.35$^{+0.09}_{-0.10}$ & --0.87$^{+0.07}_{-0.07}$ & 3.52$^{+0.24}_{-0.23}$ &0.02 \\
J1428+1653 & 16.60 & 1.04 & 9.56 & 136$^{+11}_{-14}$ & 81$^{+10}_{-12}$ & 20.61$^{+0.04}_{-0.04}$ & --0.59$^{+0.30}_{-0.29}$ & --0.09$^{+0.03}_{-0.04}$ & 6.29$^{+0.15}_{-0.25}$ &0.75 \\
J1429+0643 & 26.30 & 0.44 & 8.79 & 238$^{+85}_{-13}$ & 151$^{+30}_{-21}$ & 20.74$^{+0.05}_{-0.04}$ & --0.50$^{+0.28}_{-0.28}$ & --0.33$^{+0.04}_{-0.08}$ & 5.04$^{+0.26}_{-0.43}$ &0.78 \\
J1448\textminus{}0110 & 2.45 & 0.12 & 7.61 & 132$^{+19}_{-13}$ & 84$^{+15}_{-12}$ & 20.36$^{+0.03}_{-0.03}$ & 0.14$^{+0.19}_{-0.30}$ & --0.50$^{+0.11}_{-0.07}$ & 5.66$^{+0.27}_{-0.41}$ &1.47 \\
J1521+0759 & 8.91 & 0.48 & 9.00 & 139$^{+17}_{-24}$ & 67$^{+16}_{-22}$ & 20.33$^{+0.03}_{-0.03}$ & --0.36$^{+0.28}_{-0.34}$ & --0.31$^{+0.08}_{-0.11}$ & 6.23$^{+0.20}_{-0.46}$ &0.69 \\
J2103\textminus{}0728 & 19.50 & 0.24 & 10.28 & 200$^{+24}_{-21}$ & 72$^{+30}_{-15}$ & 20.08$^{+0.04}_{-0.04}$ & --0.03$^{+0.07}_{-0.31}$ & --1.05$^{+0.10}_{-0.11}$ & 5.43$^{+0.36}_{-0.36}$ &0.04 \\
\enddata
\tablecomments{We show galaxy properties in columns (2) -- (4) and the galactic wind and outflow properties in columns (5) -- (11). Among them, columns (2) -- (7) are fixed and adopted to constrain the model parameters in columns (8) -- (10). The galaxy properties are measured from HST/COS spectra and images as described in \cite{Xu22a}. The initial parameters (i.e., at $r$ = \Rhalf) for wind and outflow properties from the best-fit \citetalias{Fielding22} model (see Section \ref{sec:param}). We vary three parameters (columns (8) -- (10)) to fit the observed three moments of the \dNhdv\ profiles ((5) -- (7)). The 1$\sigma$ error bars are shown with the values, which are derived from the MCMC chain. The final column shows the reduced $\chi^{2}$ values of the fit.}

\end{deluxetable}

\section{Discussion} 
\label{sec:discuss}

\subsection{Comparisons to other Multiphase, Multiscale Studies of Galactic Winds}
\label{sec:comparison}

Galactic winds are of great interest because they are a primary mechanism driving feedback that shapes galaxy evolution. 
Over the last decade, significant progress has been made in understanding both the observations and the multiphase physics of these outflows.
These include multi-wavelength observational studies of galactic winds, as well as analytic and hydrodynamic models spanning a wide range of resolutions and domain sizes \citep[see reviews in][]{Veilleux20, Thompson24}.

Specifically, the prototypical starburst galaxy M 82 (SFR = 8 \Msun\ yr$^{-1}$, \Mstar\ = 10$^{10}$ \Msun, and \Rhalf\ = 300 pc) has accumulated extensive studies, owing to its proximity and well-defined biconical outflows \citep[e.g.,][]{Shopbell98, Strickland09, Leroy15, Martini18}.
In observations, we have measured the radial distribution of cool outflow-cloud properties using Subaru/FOCAS \citep{Xu23c}, and overlaid our results as black lines with gray bands (1$\sigma$ error) in Figures~\ref{fig:spatial_dist_vel}--\ref{fig:spatial_dist_Etot}.
First of all, the observed \Vcloud\ in M82 follows our model trends and closely tracks one of our galaxies, J1112+5503, which is a also massive, highly star-forming system (SFR $\sim$ 40 \Msun\ yr$^{-1}$ and \Mstar\ = 10$^{9.6}$ \Msun).
For the mass- and energy-loading factors (Figures~\ref{fig:spatial_dist_Xdot} and \ref{fig:spatial_dist_Etot}), the M~82 measurements are consistent with either flat radial evolution or with declines, while our models show similar behavior for massive galaxies.

Furthermore, \citet{Lopez25} study the properties of cool outflow clouds in M~82 from \ha\ emitting structures using HST/Advanced Camera for Surveys (ACS) imaging from \citet{Mutchler07}. They report that individual cool clouds commonly exhibit elongated elliptical or arc-like morphologies rather than the cometary structures predicted in some simulations \citep[e.g.,][]{Gronke18, Schneider20, Fielding22}. Their inferred cloud column density (\Nhcloud) is $\sim$ 10$^{20}$ -- 10$^{21}$ cm$^{-2}$, matching well with our galaxies (column 7 in Table \ref{tab:obs}). Their cloud masses, \Mcloud, lie in the range $10^{4}$--$10^{5}$ \Msun, consistent with the lower bounds in our best-fit models (column 10). Their resolved clouds reside at radii of $\sim0.5$--$2$ kpc ($\sim2$--$7 \times$\Rhalf).

Observational studies of the hot and very hot phases of galactic winds are much rarer, typically detected via extended, diffuse X-ray emission \citep[e.g.,][]{Strickland09, Liu12, Zhang14, Lopez20, Iwasawa23}.
In M~82’s central starburst ($r\lesssim$ \Rhalf), hard X-ray spectra are dominated by He- and H-like Fe at $T\sim10^{8}$~K.
The inferred terminal velocity is 1400--2200~km~s$^{-1}$ and the hot-phase mass loading is $\sim0.2$--$0.6$ \citep[e.g.,][]{Strickland09}. These values again are consistent with our modeled results for massive galaxies (e.g., right panel of Figure \ref{fig:spatial_dist_Xdot}).
In the soft X-ray, \citet{Boettcher24} analyze deep archival Reflection Grating Spectrometer (RGS) observations from \textit{XMM-Newton} and detect clear emission from \oviii\ (0.65, 0.77~keV), \nex\ (1.02~keV), and \mgxii\ (1.47~keV).
They infer a hot wind phase with $T\sim3\times10^{6}$~K and velocity $\gtrsim2000$~km~s$^{-1}$ (from \oviii), consistent with the hard X-ray estimates and with model predictions for massive galaxies in our sample.

Several simulations directly model properties of the hot wind.
\citet{Schneider20} present high-resolution ($<5$~pc) simulations of a multiphase outflow from a disk galaxy with a starburst radius of 1000~pc, \Mstar\ = 10$^{10}$ \Msun, and SFR = 5--20 \Msun\ yr$^{-1}$. 
They highlight that mixing between hot (in their case: $T>5\times10^{5}$~K) and cool ($T<2\times10^{4}$~K) gas efficiently transfers momentum at all radii.
In their high-SFR state (SFR $=20$~\Msun~yr$^{-1}$) at 35~Myr, \etamhot\ increases from 0.04 to 0.06 between 2--8~\Rhalf, while \etamcool\ decreases from 0.04 to 0.01 over the same range.
These values are lower than the ones from most of our galaxies, but they approach the lower bounds in Figure~\ref{fig:spatial_dist_Xdot}, where we indeed find cases with decreasing \etamcool, increasing \etamhot, and both values $<0.1$.
This likely reflects a regime in which the hot wind fails to launch strong cool outflows.

By contrast, other findings predict consistent hot-phase properties with most of our galaxies. For example, \citet{Li25} simulate a series of starbursts with SFR = 1.5--30 \Msun\ yr$^{-1}$, \Mstar\ = 1.6--3.3 $\times$ 10$^{8}$ \Msun, and the majority of stars forming within the
central 1000 pc. They find \etamhot\ $\sim$ 0.2--0.4 at 10~\Rhalf\ and hot-wind velocities of 800 -- 1400~km~s$^{-1}$. In addition, \citet{Nguyen21} develop an analytic hot-wind model that includes mass loading and non-spherical divergence, calibrate it to simulations, and apply it to M~82. They assume SFR = 10 \Msun\ yr$^{-1}$ and \Rhalf\ = 300 pc. They find a total mass-loading factor about 1 and predict an asymptotic hot-wind velocity of $\sim1000$~km~s$^{-1}$. They also emphasize that geometry and mass loading can commonly lower \Vwind\ and \etamhot\ relative to un–mass-loaded spherical expectations. 

Overall, our work combines observations and models of the cool and hot phases across a wide range of galaxy properties, and the resulting trends agree with prior multiphase, multiscale studies for M 82.

\subsection{Caveats}
\label{sec:caveats}

Our current methodology and the model have several caveats and some of them are tightly related to our future directions.
First of all, the model assumes a single initial cloud mass, \Mcloudinit, for all cool outflow clouds; this simplification precludes recovering a realistic cloud–mass distribution.
Then the energy-loading factors (\etaE)—particularly for the hot phase—are not included as free parameters in the present implementation.
In the CC85 framework \citep{Chevalier85}, hot-phase properties scale with the ratio $\eta_\text{E}/\eta_\text{M}$, so allowing \etaE\ to vary is essential for completeness.
In the follow-up Paper~II, we will modify the methodology and model hierarchy to address these two points, which will help us understand why several galaxies are not well fit under the current settings.

Another longstanding question is what is the best way to extract the hydrogen column density profile (i.e., \dNhdv) from rest-UV absorption lines. We adopt the \citetalias{Fielding22} model to fit \dNhdv\ derived using the partial coverage method (PCM) from CLASSY data \citep{Xu22a}. More recent studies have shown that PCM likely yields underestimated column densities. For example, \cite{Huberty24, Carr25} suggest that the underestimation is due to the finite instrumental resolution and the resulting smoothed spectral data. On the contrary, \cite{Jennings25} suggest that such underestimation is potentially caused by missed high-column-density sight lines that are optically thick which are invisible in UV spectra. While the exact biases introduced by PCM—if any—remain under active investigation, the methods presented in this paper are applicable to any \dNhdv\ distributions.

Finally, the current \citetalias{Fielding22} model does not account for the CGM component. As the galactic wind expands, it can sweep up substantial amounts of ambient CGM material, potentially altering the mass-loading factors. Future comparisons with the results from cosmological simulations, such as FIRE-2 \citep{Hopkins17}, will help assess the magnitude of this effect.

\section{Conclusion} 
\label{sec:conclusion}

In this paper, we present a method to fit unresolved rest-UV spectra of galactic outflows with multiphase, multiscale galactic-wind models.
We select 22 galaxies from the CLASSY survey with robust cool-phase outflow measurements and fit them using the analytic wind model of \cite{Fielding22}.
Our main results are summarized as follows:

\begin{enumerate}

    \item A key observable controlling the derived cool-phase outflow rates and loading factors is the hydrogen column density as a function of velocity.
    We fit three moments of it that together capture its shape: the integrated hydrogen column (\Nhcloud), the bulk velocity (\Vcloud), and the half-width at half maximum (\HWHMcloud).

    \item We identify three free parameters to fit the observables: the initial hot-phase mass loading factor (\etamhotinit), the initial cool-phase mass loading factor (\etamcoolinit), and the initial cloud mass (\Mcloudinit).
    These quantities vary with radius as the hot wind and cool clouds exchange mass, momentum, and energy.
    Other galaxy properties (e.g., stellar mass and SFR) are fixed to their observed values for each galaxy.

    \item We find 18 out of 22 galaxies are fitted well (\chisq $<$ 1.5). Among the three variables, \etamcoolinit\ is tightly constrained, with uncertainties $\lesssim0.1$~dex.
    \Mcloudinit\ and \etamhotinit\ are moderately constrained, with median uncertainties between $\sim0.2$ and 0.3~dex. These parameters should be interpreted as effective, model-conditional results that are consistent with observations.

    \item $\sim$60\% of galaxies prefer \etamhotinit\ $<$ 1.0 (peaking near 0.4), and $\sim$80\% prefer \etamcoolinit\ $<$ 1.0 (peaking near 0.1).
    These distributions are broadly consistent with state-of-the-art simulations of starburst-driven winds.
    We also find high \Mcloudinit\ is favored for our galaxies, though we caution that the current model assumes that all outflow clouds have the same mass, which is likely unrealistic and will be relaxed in our subsequent paper.

    \item The models also reveal radial profiles of key outflow properties.
    In each galaxy, the cool-phase velocity \Vcloud$(r)$ increases rapidly around $r = 1$–$2$\Rhalf, then reaches a plateau. The hot-wind velocity \Vwind$(r)$, by contrast, tends to decrease overall. Similarly, we find that the mass and energy loading factors of the cool phase commonly increase with radius, while those of the hot phase decrease. These patterns are consistent with acceleration of the cool phase by the hot wind through mass, momentum, and energy exchange. Combining both phases, the total energy flux drops by 10–90\% over radius, likely radiated away via the turbulent radiative mixing layer between the cool and hot phases.
    
    \item Comparisons with observational and simulation studies of multiphase winds show broad consistency in the cool and hot phase velocities, mass/energy loadings, and their radial trends.

\end{enumerate}
In sum, this paper adopts the default \citetalias{Fielding22} configuration to validate two key ideas: velocity–radius mapping from \dNhdv\ enables spatially integrated spectra to constrain spatially resolved structure, and cool–hot phase coupling can allow indirect constraints on the hot phase without direct detections.

There are several promising directions for expanding current frameworks in future studies. First, galactic outflows leave imprints not only in UV absorption lines but also in optical emission lines \citep[e.g.,][]{Wood15, Freeman19, Weldon24}. Combining both diagnostics has been shown to more effectively constrain outflow models \citep[e.g.,][]{Carr18, Carr23, Peng24, Xu25} and should be considered in future studies. Second, none of our galaxies currently exhibits a direct detection of the hot phase due to the absence of high-resolution X-ray spectroscopy, which hinders one-to-one comparisons with model predictions. This limitation may be addressed with data from current and upcoming X-ray observatories, such as \textit{XRISM} \citep{XRISM20}, \textit{Lynx} \citep{Gaskin19}, and \textit{Athena} \citep{Barret20}. A few of our galaxies (e.g., J0036--3333, also known as Haro 11; \citealt{Menacho19}) have spatially resolved spectra for ionized outflows. Acquiring more such data—using integral field units like VLT/MUSE and JWST/NIRSpec—and directly comparing the resolved observations with spatially resolved models would provide valuable insights.

\begin{acknowledgments}
X.X. acknowledges the fellowship funding from Center for Interdisciplinary Exploration and Research in Astrophysics (CIERA), Northwestern University. D.B.F. gratefully acknowledges support from NSF through grants AST-2407387 and from NASA through grants HST-AR-17859.015-A and HST-AR-17559.009-A.

The CLASSY team is grateful for the support for program
HST-GO-15840, which was provided by NASA through a
grant from the Space Telescope Science Institute, which is
operated by the Associations of Universities for Research in
Astronomy, Incorporated, under NASA contract NAS5-26555.
B.L.J. acknowledges support from the European Space Agency
(ESA). C.L.M. gratefully acknowledges support from NSF
AST-1817125. The CLASSY collaboration extends special
gratitude to the Lorentz Center for useful discussions during the
“Characterizing Galaxies with Spectroscopy with a view for
JWST” 2017 workshop that led to the formation of the
CLASSY collaboration and survey.

CAFG was supported by NSF through grants AST-2108230 and AST-2307327; by NASA through grants 21-ATP21-0036 and 23-ATP23-0008; and by STScI through grant JWST-AR-03252.001-A.

GLB acknowledges support from the NSF (AST-2108470, AST-2307419), NASA TCAN award 80NSSC21K1053, and the Simons Foundation

\end{acknowledgments}

\facilities{HST(COS)}

\software{astropy \citep{2013A&A...558A..33A,2018AJ....156..123A,2022ApJ...935..167A},  
          Cloudy \citep{2013RMxAA..49..137F}, ChatGPT (OpenAI, 2023; \url{https://chat.openai.com/chat})
          }

\bibliography{sample7}{}
\bibliographystyle{aasjournal} 

\end{document}